\documentclass[12pt]{article}

\usepackage{amsmath}
\usepackage{amssymb}
\usepackage{amsthm}

\usepackage{array}
\usepackage{multirow}
\usepackage{booktabs}

\usepackage{url}
\usepackage{graphicx}
\graphicspath{{FIGS/}}

\usepackage{color}
\usepackage{times}

\usepackage[latin1]{inputenc}

\usepackage[a4paper,text={6.5in,9.0in},headheight=1in,bottom=1in,hcentering]
           {geometry}

\DeclareMathOperator*{\argmax}{arg\,max}

\DeclareMathOperator*{\eqdef}{\stackrel{\text{\tiny def}}{=}}

\def\RR{R}

\def\yes{\mathrm{yes}}
\def\no{\mathrm{not}}

\def\QQ{Q}
\def\Q{q}

\def\WW{W}

\def\XX{X}
\def\X{\mathcal{X}}
\def\x{\mathrm{x}}

\def\LL{L}

\def\BB{B}
\def\B{b}

\def\S{s}

\definecolor{grey}{rgb}{0.5,0.5,0.5}
\definecolor{darkred}{rgb}{0.6,0.1,0.1}

      \def\vppp{\vspace{3mm}}
       \def\vp{\vspace{1mm}}
    \def\vnnn{\vspace{-3mm}}
\def\vnnnn{\vspace{-4mm}} \def\vnn{\vspace{-2mm}}    \def\vn{\vspace{-1mm}}

\begin{document}

\title{A Probabilistic Framework for Lexicon-based\\
       Keyword Spotting in Handwritten Text Images}
\author{E. Vidal, A.H. Toselli, J. Puigcerver} 
%
\maketitle

\begin{abstract}
  Query by String Keyword Spotting (KWS) is here considered as a key
  technology for indexing large collections of handwritten text images
  to allow fast textual access to the contents of these colections.
  Under this prespective, a probabilistic framework for lexicon-based
  KWS in text images is presented.
  The presentation aims at providing a tutorial view which helps
  understanding the relations between classical statements of KWS and
  the relative challenges entailed by these statements. 
  More specifically, the development of the proposed framework makes
  it self-evident that word recognition or classification 
  implicitly or explicitly underlies any formulation of KWS.
  Moreover, it clearly suggests that the same statistical models and
  training methods successfully used for handwriting text recognition,
  can advantageously used also for KWS, even though KWS does not
  generally require or rely on any kind of previously produced image
  transcripts.
  These ideas are developped into a specific, probabilistically sound
  approach for segmentation-free, lexicon-based, query-by-string KWS.
  Experiments carried out using this approach are presented, which
  support the consistency and general interest of the proposed
  framework.
  Several datasets, traditionally used for KWS benchmarking are
  considered, with results significantly better than those previously
  published for these datasets.
  In addition, results on two new, larger handwritten text image
  datasets are reported, showing the great potential of the methods
  proposed in this paper for indexing and textual search in large
  collections of handwritent documents.
\end{abstract}


                                \newpage

\section{Introduction}                                \label{sec:intro}


Massive quantities of historical manuscripts have been converted into
high resolution images in the last decades as a result of
digitalization works carried out by archives and libraries world wide.
Billions of handwritten text images have been produced through these
efforts, and this is only a minuscule part of the amount of
handwritten documents which are still waiting to be digitalized.
The aim of manuscript digitization is not only to improve
preservation, but also to make the handwritten documents easily
accessible to interested scholars and general public.
However, access to the real wealth of these images, namely, their
\emph{textual contents}, remains totally elusive.
Consequently, there is a fast growing interest in automated methods
which allow the users to search for relevant textual information
contained in handwritten text images.

%

In order to use classical Information Retrieval (IR)
methods~\cite{Manning:2008} for plain-text indexing and search, a
first step would be to convert the handwritten text images into
digital text.
But the image collections for which text indexing is highly in demand
are so large that the cost of manually transcribing these images is
entirely prohibitive, even using crowd-sourcing approaches.
An obvious alternative to manual transcription is to rely on
automatic Handwritten Text Recognition
(HTR)~\cite{Bazzi99,Bunke04,Toselli04}.  However, despite the great
recent advances in the field~\cite{Graves:09,Espana:11,Sanchez:16},
fully automatic transcripts of the kind of historical images of
interest still lack the accuracy required to enable useful plain-text
indexing and search.
Another possibility is to use computer-assisted transcription
methods~\cite{toselli10a,Romero12a}, but so far these methods can
not provide the huge human-effort reductions needed to render
semiautomatic transcription of large image collections
feasible~\cite{toselli17}.


HTR accuracy becomes low on real historical handwritten text images
for many reasons, including unpredictable, erratic layouts, lines with
uneven interline spacing and highly variable skew, etc.  In addition,
unambiguous reading order of layout elements is often difficult or
impossible to determine.
Current state of the art HTR systems achieve good transcription
results only if perfect layout, line detection and reading order are
taken for granted (as it is often the case in published results).
Clearly, for moderate sized image collections, many of these problems
can often be fixed by simple and inexpensive manual post-processing,
but this is completely impracticable when collections of hundreds of
thousands, or millions of images are considered.


Interestingly, most or all of these problems disappear or become much
less severe if, rather than to achieve accurate word-by-word image
transcripts, the goal is to determine how likely is that a given word
is or is not written in some indexable image region, such as a text
line, an arbitrary text block, or just a full page image.
This goal statement places our textual information retrieval problem
in the field broadly known as Keyword Spotting (KWS).
A comprehensive survey of KWS techniques has recently been published
in~\cite{Giotis:2017}.
%
Among the works cited in this survey, it is worth noting that many
recent developments are inspired in one form or the other in earlier
KWS works in the field of automatic speech recognition (ASR), such
as~\cite{Christiansen:77,Rose:90,Szoke:05,Chelba:06,Chia:10}.  This is
also the case of the work presented in this paper.


Generally speaking, KWS aims at determining \emph{locations} on a text
image or image collection which are likely to contain instances of the
query words, without explicitly transcribing the image(s).
One of the earliest statement of KWS for text images, known as
\emph{``segmentation-based''}~\cite{Giotis:2017}, assumes the
\emph{locations} to be previously cropped small image regions which
contain individual words.  Adopting word-sized image regions is a very
tempting assumption.  But manual word segmentation is obviously
unfeasible for large image collections, and automatic word
segmentation can not be reliably carried out, because word separation
is often nonexistent and/or notoriously inconsistent in most
historical handwritten documents.
%
%
This is in contrast with \emph{``segmentation-free''} KWS
formulations~\cite{Giotis:2017}, which consider that finding the word
locations and determining how likely the corresponding image regions
may contain a query word are dual aspects of the same problem.

Many recent works assume an intermediate view of KWS were relatively
large image regions (such as lines or paragraphs), which typically
contain several words, are considered the search targets where word
relevance likelihoods have to be determined.
%
This view, which is often referred to also as
\mbox{(word-)segmentation-free}, is called
\emph{``line-segmentation-based''} in~\cite{Giotis:2017}.  It is
particularly interesting because it can very adequately support the
kind of indexing and search features needed to provide textual access
to large collections of handwritten images -- and, moreover, automatic
image segmentation into these larger regions is generally very much
less problematic than individual word segmentation.
Most of the developments and results of this paper loosely adhere to
this view.


Another traditional taxonomy in KWS distinguishes
\emph{Query-by-Example} (QbE) and \emph{Query-by-String} (QbS)
formulations, depending on whether query words are specified by means
of example-images or just as character strings,
respectively~\cite{Giotis:2017}.  While there are applications for
which the QbE scenario can certainly be useful, it is clearly
inappropriate for effective indexing and search in large image
collections.  Therefore, in this paper only the QbS framework is
considered.


As we will see, HTR and KWS can advantageously share the same
statistical models and training methods.  However, it is important to
realize that HTR and KWS are fundamentally different problems, even if
both may rely on identical probability distributions and models.
The HTR decision rule attempts to obtain the best sequence of words
(transcript) for a given (line) image region.  Therefore the result
epitomizes just the \emph{mode of the distribution}; once a transcript
has been obtained, the distribution itself can be safely discarded.
In contrast, KWS decisions are delayed to the query phase and, for each
decision, the \emph{full distribution} can (should!) be used.
%
This obviously explains why proper KWS can always achieve better
overall indexing and search results than those provided by naive KWS
based on plain HTR transcripts.

%

An indexing and search system can be evaluated by measuring its
\emph{precision} and \emph{recall} performance for a given (large) set
of keywords. Precision is high if most of the retrieved results are
correct while recall is high if most of the existing correct results
are retrieved.
In the case of indexed automatic HTR transcripts, precision and recall
are fixed numbers, which are obviously closely correlated with the
accuracy of the recognized transcripts.
In contrast, for a KWS system based on the likelihood that a keyword
is written in an image region, arbitrary precision-recall tradeoffs
can be obtained by setting a threshold to decide whether the
likelihood is high enough or not.
We refer to this flexible search and retrieval framework as the
\emph{``precision-recall tradeoff model''}.
Under this model, it becomes even more clear that proper KWS has the
opportunity of achieving better results than naive KWS based on HTR
transcripts, as previously discussed.




Some of the developments and results presented in this paper are based
on techniques described in~\cite{toselli16}, or follow research
directions outlined in that paper.
Contributions of this paper to the state of the art in KWS for
handwritten image indexing and search include:
First, a sound probabilistic framework is presented which helps
understanding the relations between classical statements of KWS and
the relative challenges entailed by these statements.
Second, the development of this framework makes it self-evident that
word recognition implicitly or explicitly underlies any formulation
of KWS, and clearly suggests that the same statistical models and
training methods successfully used for HTR can advantageously used
also for KWS.
Third, these ideas are developed into a specific, probabilistically
sound approach for segmentation-free, lexicon-based, query-by-string
KWS.
Fourth, experiments carried out using this approach on datasets
traditionally used for KWS benchmarking yield results significantly
better than those previously published for these datasets.
And fifth, KWS results on two new, larger handwritten text image
datasets are reported, showing the great potential of the methods
proposed in this paper for accurate indexing and textual search in
large collections of handwritten documents.


The remaining sections of this paper are as follows:
The proposed general framework is introduced in
Sec.\,\ref{sec:probFramework} and developed in the following sections.
Sec.\,\ref{sec:pgram} introduces the concept of pixel-level word
posteriors.  While this concept is instructive, the computational
costs entailed are exceedingly high.  Therefore,
in Sec.\,\ref{sec:regionKWS}, we develop the idea of computing
relevance probabilities for adequate sized image regions
%
and explain how these relevance probabilities can be accurately and
efficiently computed when the image regions considered are text line
regions.
In Sec.\,\ref{sec:approaches} we briefly review popular KWS approaches
under the proposed statistical framework and discuss our specific
proposal.
The experimental settings are presented in Sec.\,\ref{sec:experiments}
and the corresponding results in Sec.\,\ref{sec:results}.
Finally Sec.\,\ref{sec:conclusion} concludes the paper summarizing the
work carried out and outlying future avenues of research.

\section{A Probabilistic Framework for Word Spotting}
                                                 \label{sec:probFramework}

In the literature of Key Word Spotting for Text Images, outlined in
Sec.\,\ref{sec:intro}, two main questions are being considered,
regarding a query word $v$ and a certain text image or image region
$\X$:
%
\begin{enumerate}\itemsep=0em
\item Is the word $v$ written in $\X$?
\item What are the locations (if any) of word $v$ within $\X$?
\end{enumerate}

The first question is a ``simple'' yes/no question which, from a
probabilistic point of view, must be somehow modeled by a binary
random variable.
%
%
The second question may appear more complex, but it can be
reformulated by asking whether each individual location within $\X$,
may contains $v$ or not.  Each of these individual questions is again
a ``simple'' yes/no question which can be modeled by a binary random
variable, as well.
%
The probabilistic framework presented here deals with these questions.

First, we introduce a random variable $\XX$ over the set of all image
regions considered.
%
A value of this random variable (i.e., an arbitrary image region),
will be denoted as $\X$.
%
%
At this point we do not need to consider what are the possible sizes
and shapes of image regions (a page, a paragraph, a line, etc.) or how
they are represented (direct pixel values, or features extracted from
these values).  Therefore, until we need to be more specific, we will
simply use the term \emph{``image''} for a value of $\XX$.
%

Second, we introduce another random variable, $\QQ$, over the set of
all possible user queries.  An arbitrary value of this random variable
will be generally denoted as $\Q$.
To keep the presentation simple, in this paper we will consider only
single-word queries; that is, we
%
focus on conventional word-based QbS KWS, where queries are individual
words $v$ from a given vocabulary $V$, which consists of a set of
words we are interested in%
\footnote{A ``non-word'' token may also be included in $V$ to account
  for image regions without text.%
}.
Nevertheless, the proposed probabilistic framework can properly
accommodate arbitrary types of queries: from single words, to regular
expressions of characters or words, or even ``example image patches'',
as in QbE KWS (see, for instance~\cite{Noya:17,vidal:15}).

Third, a random variable over the set of all possible
\emph{locations} is needed.  Ideally, one would like to consider a
word \emph{``location''} as the set of pixels that make up the image
rendering of this word.
However, such a fine-grained assumption is not generally feasible, nor
it is needed in practice, and the location of a word is often
considered to be its \emph{``bounding box''}.
%
%
Therefore, we define the random variable $\BB$ over the set of all
word-sized bounding-boxes in an image and we will use $\B$ to denote a
specific word location or bounding box.

Finally, we need the binary random variable that we referred to in the
very beginning of this section to model the event that a certain image
$\X$ (or a particular location, $\B$ within it) contains (or not) a
specific query $\Q$.  It will be named $\RR$, after
\emph{``relevant''}, in order to follow common notation in the field
of IR.  This entails a reformulation of the original question as: ``is
the image $\X$ \emph{relevant} for the query $\Q$?'', considering that
$\X$ is relevant for $\Q$ if one (or more) instance(s) of $\Q$ are
rendered in $\X$.
%

Using these random variables, we introduce two probability distributions
that
will help answer the original questions posed by KWS:
%
\begin{eqnarray}
P(\RR = \yes \!&\! \mid \!&\! \XX=\X,\,\QQ=\Q) \label{eq:def1}\\[-0.2em]
P(\RR = \yes \!&\! \mid \!&\! \XX=\X,\,\QQ=\Q,\,\BB=\B) \label{eq:def2}
\end{eqnarray}
The first distribution is the probability that the image $\X$ is
relevant for the query $\Q$, while the second is the probability that
the location $\B$ within $\X$ is relevant for the query $\Q$.
It is obvious to see that the relevance probabilities defined by
Eq.\,\eqref{eq:def1} and Eq.\,\eqref{eq:def2} can be properly
interpreted as the statistical expectation that $\Q$ is written in
$\X$ or in~$b$, respectively.

To simplify notation, in what follows we will write $P(R\cdots)$
rather than $P(R=\yes~\cdots)$, except when the full notation helps
avoiding ambiguity and/or enhancing clarity.  Similarly, other random
variables will be omitted; i.e., we will write $P(a\,\cdots)$ rather
than $P(A\!=\!a\,\cdots)$.

Simple \emph{marginalization} can be applied to derive the first, more
coarsely-grained distribution from the second one:
\begin{eqnarray}\label{eq:basicMarginal}
    P(\RR\mid\X,\Q) 
    ~=~ \sum_{\B\in\BB}P(\RR,\B\mid\X,\Q)
    ~=~ \sum_{\B\in\BB}P(\RR\mid\X,\Q,\B)~P(\B\mid\X,\Q)  
\end{eqnarray}

In the next sections we explain how to compute relevance distributions
for given images.
%
We anticipate that most of these developments will relay on the
application of the same fundamental marginalization rule used in
Eq.\,\eqref{eq:basicMarginal}.
We will start in Sec.\,\ref{sec:pgram} by introducing the concept
of word posteriors computed at the pixel level.  While such a
representation is conceptually enlightening, its computation is
expensive and, moreover, it would require prohibitive amounts of
memory and time for keyword indexing and search.  Therefore we will
argue that keyword search does not really need such a fine-grained
resolution and, in Sec.\,\ref{sec:regionKWS}, we discuss the
convenience of computing the required probabilities for whole image
regions of adequate size.  Then, in Sec.\,\ref{sec:regionYNprobs}
and\,\ref{sec:YNfromPgram} we explain how $P(\RR\mid\X,\Q)$ can be
derived from pixel-level word posteriors when $\X$ is an adequate
image region.  Finally, in Sec.\,\ref{sec:lineKWS} we explain how
these region-level word relevance probabilities can be accurately and
efficiently computed when the image regions considered are text line
regions.

\section{Pixel Level Keyword Search: Image Posteriorgram}
                                                         \label{sec:pgram}
\vnn

We assume that each handwritten page image%
\footnote{The term ``page image'' is used here for the result of
 scanning a significant piece of (handwritten) document.
} %
has undergone basic preprocessing steps including correction of
overall page skew and other simple geometrical
distortions~\cite{Kavallieratou:06,Mpastorg04a,Sauvola:00,Sepideh:13}.
We do \emph{not} need to assume that preprocessing includes any kind
of character or word segmentation.  Line segmentation is not
essentially needed either.  Nevertheless, as discussed later on, for
effectiveness, efficiency and simplicity, text can be assumed to be
organized into distinguishable, roughly horizontal lines.


The \emph{posteriorgram} of a text image $\X$, is the probability that
the query word $\Q=v\in V$ uniquely and completely appears in a
bounding box containing the pixel
$(i,j)$. 
In mathematical notation:
\begin{equation}\label{pgram0}
  P(\QQ=v\mid \XX=\X, \LL=(i,j)) ~~\equiv~~
  P(v\mid \X,i,j),~~ 1\leq i\leq I,~ 1\leq j\leq J,~~ v\in V
\end{equation}
where $\LL$ is a random variable over the set of locations (pixel
coordinates) and $I, J$ are the horizontal and vertical dimensions of
$\X$.  $P(v\mid \X,i,j)$ is a proper probability distribution over the
vocabulary $V$; that is:
\begin{equation}
  \sum_{v\in V}\,P(v\mid \X,i,j) = 1,~~ 1\leq i\leq I, 1\leq j\leq J
\end{equation}

\begin{figure}[htb]
\centering
\hspace{-0.5em}\includegraphics[width=0.4\linewidth]{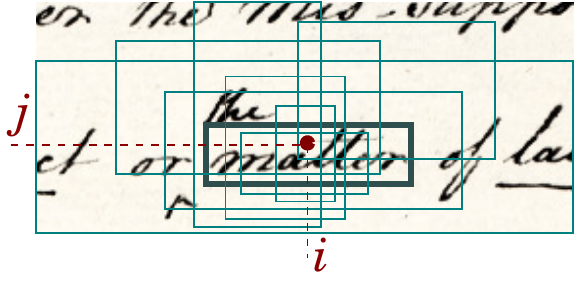}
\vnnn
\caption{\small \label{fig:marginal} %
  Bounding boxes $\B\in\mathcal{B}(i,j)$.  For
  $v=$~''\texttt{\textbf{matter}}'', the thick-line box will provide
  the highest value of $P(v\mid \X,\B)$, while most of the other boxes
  will contribute only (very) low values to the sum.%
}
\end{figure}

\begin{figure}[tb]
\centering
\includegraphics[width=0.8\textwidth]{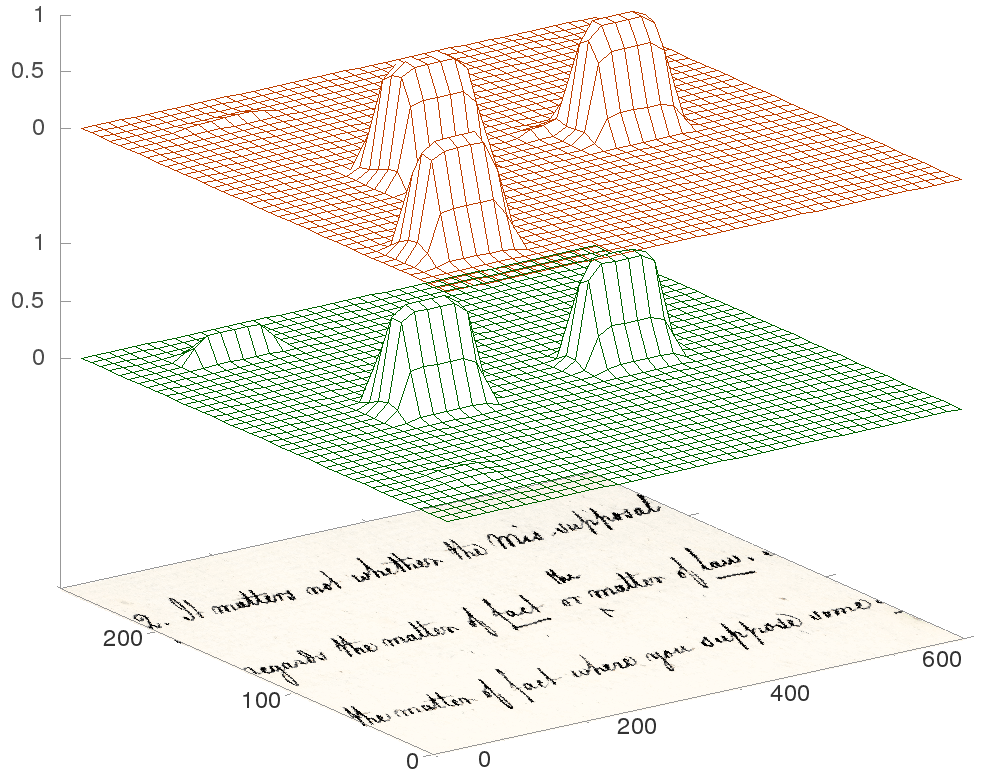}
\hspace{-6em}%
\begin{minipage}[b]{.2\textwidth}
  {\small $P_2(v\mid \X,i,j)$}

  \vspace{6.3em}
  {\small $P_0(v\mid \X,i,j)$}

  \vspace{6.1em}
  {\small $\X$}

  \vspace{7.5em}
\end{minipage}

\vnnn
\caption{\small                                  \label{fig:2Dpostgm} %
  Well trained optical HMM classifiers were used to compute
  two 2-D posteriorgrams for a text image $\X$ and keyword
  $v=$``\textbf{\texttt{matter}}''.  $P_2$ was obtained using a
  contextual, bigram-based classifier, and provides much better
  posterior estimates than those of $P_0$, computed using a
  context-agnostic isolated word classifier based on identical optical
  models.%
}
\vnn
\end{figure}

A simple way to compute $P(v \mid \X,i,j)$ is by considering that $v$
may have been written in any possible bounding box $\B$ of $\X$.
%
%
\begin{equation}\label{eq:pgram0a}
  P(v\mid \X,i,j) ~=
  \sum_{\B\in\mathcal{B}(i,j)}\!P(v,\B \mid \X , i, j) ~=
  \sum_{\B\in\mathcal{B}(i,j)}\!P(\B\mid \X,i,j) P(v\mid \X,\B,i,j)
\end{equation}
where $\mathcal{B}(i,j)$ is the set of all bounding boxes of the image
which contain the pixel $(i,j)$.  Fig.\,\ref{fig:marginal} illustrates
this marginalization process.
%
%
$P(v \mid \X, \B, i, j)$ is the probability that $v$ is the (unique)
word written in the box $\B$ (which includes the pixel $(i, j)$).
Therefore it is conditionally independent of $(i,j)$ given $\B$,
and Eq\,\eqref{eq:pgram0a} simplifies to:
%
\begin{equation}\label{eq:pgram0b}
P(v \mid \X, i, j) =
\sum_{\B \in \mathcal{B}(i,j)} P(\B \mid \X, i, j) P(v \mid \X, \B)
\end{equation}
Real results of computing the posteriorgram $P(v\mid\X,i,j)$ in this
way for a given image $\X$ and a specific keyword $v$ are shown in
Fig.\,\ref{fig:2Dpostgm}.

The distribution $P(\B \mid \X, i, j)$ of Eq.\,\eqref{eq:pgram0b}
should be interpreted as the probability that some word (not
necessarily $v$) is written in the image region delimited by the
bounding box $\B$ (which includes the pixel $(i,j)$).
Therefore, this probability should be high for word-shaped and
word-sized bounding boxes centered around the pixel $(i,j)$, like some
of those illustrated in Fig.\,\ref{fig:marginal}.  In contrast, it
should be low for boxes which are too small, too large, or are too
off-center with respect to $(i,j)$.  For simplicity, we could assume
that this distribution is uniform for all reasonably sized and shaped
boxes around $(i,j)$ and then just replace this distribution with a
constant in Eq.\,\eqref{eq:pgram0b}.

On the other hand, the term $P(v\!\mid\! \X,\B)$, is exactly the
probability implicitly or explicitly computed by any system capable of
recognizing a pre-segmented word image (i.e., a sub-image of $\X$
bounded by $\B$).  Actually, such an isolated word recognition task
can be formally written as the following classification problem:
\begin{equation}\label{eq:class0}
  \hat{v} ~=~ \argmax_{v\in V}~P(v\mid \X,\B)
\end{equation}
%
In general, any system capable of recognizing
pre-segmented word images implicitly or explicitly computes %
$P(v\mid \X,\B)$ and can thereby be used to obtain the posteriorgram
according to Eq.\,\eqref{eq:pgram0b}.

Obviously, the better the classifier, the better the corresponding
posteriorgram estimates.  This is illustrated in
Fig.\,\ref{fig:2Dpostgm}, which shows two examples of image
posteriorgram obtained according to Eq.\,\eqref{eq:pgram0b} using two
different word image recognizers.
In both cases, well trained optical hidden Markov models (HMM) were
used to compute $P(v\mid \X,\B)$ $\forall B\in\mathcal{B}(i,j)$.
$P_0(v\mid \X,i,j)$ was obtained directly, using a plain,
context-agnostic optical recognizer, and $P_2(v\mid \X,i,j)$ was
produced using a more precise %
\emph{contextual word recognizer}, additionally based on a well
trained bigram.
As it can be seen, $P_0$ values are only good for the two clear
instances of ``\textbf{\texttt{matter}}'', and almost vanish for a
third instance, probably because of the faint character
``\textbf{\texttt{m}}''.  Worse still, $P_0$ values are relatively
high for the similar, but wrong word ``\textbf{\texttt{matters}}''; in
fact very much higher than for the third, faint instance of the
correct one.
In contrast, the contextual recognizer led to high $P_2$ values for
all the three correct instances of ``\textbf{\texttt{matter}}'', even
for the faint one, while the values for the wrong word were very low.
Clearly bigrams such as ``\textbf{\texttt{It matter}}'' and %
``\textbf{\texttt{matter not}}'' are unlikely, thereby preventing
$P_2(v\mid \X,\B)$ to be high for any $\B$ around the word
``\textbf{\texttt{matters}}''.  On the other hand, the bigrams
``\textbf{\texttt{the matter}}'' and ``\textbf{\texttt{matter of}}''
are very likely, thereby helping the optical recognizer to boost
$P_2(v\mid \X,\B)$ for boxes $\B$ around the faint instance of
``\textbf{\texttt{matter}}''.

Pixel-level posteriorgrams could be directly used for keyword search:~
Given a threshold $\,\tau\in[0,1]$,\, a word $v\in V$ is spotted in
all image positions where $P(v\mid \X,i,j)>\tau$.  Varying $\tau$,
adequate \emph{precision}--\emph{recall} tradeoffs could be achieved.

\section{Image-Region Keyword Indexing and Search}
\label{sec:regionKWS}
\vnn

Computing the full posteriorgram as in Eq.\,\eqref{eq:pgram0b} for
all the words of a large vocabulary (as needed for indexing purposes)
and all the pixels of each page image of the collection to be indexed
entails a formidable amount of computation.  Therefore such a direct
approach becomes completely prohibitive for the size of handwritten
text image collections which are the target of the indexing methods
considered in this paper.  The same can be said for the exorbitant
amount of memory which would be needed to explicitly store all the
resulting posterior probabilities.
Therefore, rather than working at the \emph{pixel level}, some
adequate \emph{image regions}, $\x$, which are smaller than $\X$ and
are suitable search targets for users, need be defined to compute the
relevance probabilities introduced in Sec.\,\ref{sec:probFramework}.
While this problematic is seldom discussed explicitly in the
traditional KWS literature, \emph{region proposal}~\cite{Ren:17}
has been the focus of a number of studies in the object recognition
community -- see e.g.~\cite{En:16}, which deals with graphic pattern
spotting in historical documents.

In the traditional KWS literature, word-sized regions are often
considered.  This is reminiscent of segmentation-based KWS approaches
which required previously cropped accurate word bounding boxes.
However, as discussed in Sec.\,\ref{sec:intro}, this is not realistic
for large image collections.
%
%
More importantly, by considering isolated words, the underlying
word recognizer can not take advantage of word linguistic context to
achieve good spotting precision (as illustrated in
Fig.\,\ref{fig:2Dpostgm}).

At the other extreme we may consider whole page images, or relevant
\emph{text blocks} thereof, as the search target image regions.
While this can be sufficiently adequate for many indexing and search
applications, a page may typically contain many instances of the
word searched for and, on the other hand, users generally like to get
narrower responses to their queries.

                                                        \enlargethispage{1em}

A particularly interesting intermediate search target level consists
of \emph{line-shaped regions}.  Lines are useful targets for indexing
and search in practice and, in contrast with word-sized image regions,
lines generally provide sufficient linguistic context to allow
computing very precise word classification probabilities.  Moreover,
as will be discussed in Section\,\ref{sec:approaches}, line region
posteriorgrams can be very efficiently computed.

\subsection{Image-Region Word Posteriors are not Adequate}
                                                   \label{sec:regionClass}

A fairly widespread idea in recent works on line-region KWS is to
build on whole-region word posterior probabilities, $P(v\mid \x)$.
Despite their popularity, it should be pointed out that these word
posteriors are \emph{not} directly suitable for region-level keyword
search; Instead, they are appropriate to solve the following
$|V|$--class classification problem:

\medskip
\centerline{%
  ``Given an image region $\x$, find a (\emph{single}) word
    $v\in V$ which most likely appears~in~$\x$''
}
\medskip

\noindent
The associated \emph{Bayes decision rule}~\cite{Duda73} is to classify
$\x$ as:
\begin{equation}\label{eq:wordClassif}
  \hat{v} ~=~ \argmax_{v\in V} P(v\mid \x)
\end{equation}

\vnn
The intuition behind this classification problem and the corresponding
decision rule is unclear: What is the meaning of the (unique) %
``most likely word'' $\hat{v}$ we are looking for?  Are we searching
for a kind of ``most dominant'' word within $\x$?~
Moreover, $P(v\mid \x)$ sums up to one for all $v\in V$ and therefore
the decision rule in Eq.\,\eqref{eq:wordClassif} ``discriminates'' among
words in $V$; but, for keyword search, each word actually written in
$\x$ should have high probability and we should rather expect that the
sum to be much larger than $1$ -- in fact it should approach the
expected number of different words written in $\x$
(cf. Sec.\,\ref{sec:kwsHtr}).

%
The following example illustrates these inconsistencies according to
the probabilistic framework introduced in
Sec.\,\ref{sec:probFramework}.
Suppose a text image $\x$ contains the text %
``\texttt{\textbf{the cloud is white}}''.  Then, an \emph{ideal}
relevance distribution would be:
\begin{eqnarray*}
P(\RR\mid v,\x) &\approx&
\begin{cases}
1 & v = \text{``\texttt{\textbf{the}}''}\\[-0.2em]
1 & v = \text{``\texttt{\textbf{cloud}}''}\\[-0.2em]
1 & v = \text{``\texttt{\textbf{is}}''}\\[-0.2em]
1 & v = \text{``\texttt{\textbf{white}}''}\\[-0.2em]
0 & \text{otherwise}
\end{cases}
\end{eqnarray*}
In contrast, the values of $P(v\mid\x)$ should be much lower for all
the four words appearing in $\x$,
while the probabilities for other non-relevant words may be
significant.
%
%
%


It is worth noting, however, that $P(v\mid \x)$ is perfectly adequate
if $\x$ is reduced to just the bounding box of each single written
word to be searched for.  This is in fact the advantage which allows
some segmentation-based approaches to KWS to achieve reasonable results
-- at the expense of requiring previously given accurate word
segmentations of the text images.

                                                        \enlargethispage{1em}

Assuming for simplicity that the positions $(i,j)$ where the words to
be searched for may appear are uniformly distributed, $P(v\mid \x)$ can
be readily obtained as a pixel-average of the posteriorgram:
\begin{equation}\label{eq:wordpost}
  P(v\mid \x) ~= \sum_{ij} P(v, i,j\mid \x)
             ~= \sum_{ij} P(i,j\mid \x) P(v\mid \x,i,j)
             ~\approx~ \frac{1}{I\,J}\sum_{ij}P(v\mid \x,i,j)~
\end{equation}
where $IJ$ is the number of pixels of $\x$.
This approach will be used in Sec.~\ref{sec:results}, to empirically
confirm that using $P(v\mid \x)$ actually leads to poor line-region
KWS results.

\subsection{Proper Classification Model for
                  Image-Region Keyword Search} \label{sec:regionYNclass}

Clearly, the classification problem underlying region-level KWS is
\emph{not} \emph{word recognition}.  Instead, KWS entails a
related but different 2-class classification problem for each
word $v\in V$:

\vspace{1.3em}
\begin{equation}\label{eq:yndef}\end{equation}
\vspace{-5.5em}
\begin{center}
\begin{minipage}{0.8\textwidth}
Given $v$, classify each line image $x$ into one of two classes:
\begin{itemize}\vnn\itemsep=-3pt
\item $\yes\,$: $v$ is (one of the words) written (somewhere) in $\x$
\item $\no\,$:  $v$ does not appear in $\x$
\end{itemize}
\end{minipage}
\end{center}
%
%
As in any binary decision problem, the \emph{cost} of the decision
taken can be decomposed in a \emph{loss} matrix, or function:
\begin{center}
\begin{tabular}{ll|cc}
&     & \multicolumn{2}{c}{Decision} \\
&     & No & Yes \\
\hline
\parbox[t]{2mm}{\multirow{2}{*}{\rotatebox[origin=c]{90}{Truth}}}
& No  & $\lambda_{NN}$ & $\lambda_{NY}$ \\
& Yes & $\lambda_{YN}$ & $\lambda_{YY}$ \\
\end{tabular}
\end{center}
Then, using the posterior probability underlying this classification
problem, which is in fact the relevance distribution introduced in
Sec.\,\ref{sec:probFramework}, $P(\RR \mid \x,v)$, the \emph{optimal}
decision (i.e. the Bayes, or Minimum Expected Risk,
decision) \cite{Duda73}, results in answering $\yes$ \emph{iff}:
%
\begin{equation}\label{eq:YNdecision}
  P(\RR \mid \x,v) 
  ~>~ \tau ~=~
  \frac{\lambda_{YN} - \lambda_{NN}}
  {\lambda_{NY} - \lambda_{YY} + \lambda_{YN} - \lambda_{NN}}
\end{equation}
%
Note that in this two-class case, $\lambda$ reduces to a single
threshold $\tau$, which can be adjusted to achieve varying
\emph{precision}--\emph{recall} tradeoffs, as required in KWS.
A special case of this decision rule is when making no errors is not
penalized (i.e. $\lambda_{NN} = \lambda_{YY} = 0$) and all errors are
penalized equally (i.e $\lambda_{YN} = \lambda_{NY}$). In such special
case, the \emph{optimal} threshold reduces to $\tau = 0.5$.

\medskip
In the following sub-sections we will explain different ways to
properly and efficiently compute $P(\RR\mid \x,v)$.



\subsection{Computing Image-Region Relevance Probabilities}
                                                   \label{sec:regionYNprobs}

Let the correct transcript of an image region $\x$ be the sequence
of words $w = w_1, w_2,\dots,w_n$, $w_k\in V, 1\leq k\leq n$,
and let us abuse the notation and use $v\in w$ to denote that
$\exists k,~w_k = v$.
The definition of the class $\yes$ in Eq.\,\eqref{eq:yndef} can then
be written as:
%
\begin{equation}\label{eq:ynor}
  (\RR=\yes) ~\equiv~ (w_1=v\vee w_2=v \ldots \vee w_n=v)
       ~\equiv~ (v\in w)
\vn
\end{equation}

Of course, if $w$ were known, the relevance probability
$P(\RR\mid\x,v)$ would trivially be $1$ if $v\in w$ and $0$ otherwise.
In practice, no transcripts are available, but an obvious, naive idea
is to approximate $w$ with a best transcription hypothesis, $\hat{w}(\x)$,
produced by a HTR system (see Sec.\,\ref{sec:kwsHtr}):
\vn
\begin{equation}\label{eq:1bOrApprox}
P(\RR\mid \x,v) ~\approx~
  \begin{cases}
    \,1 & \text{if $\,v\in \hat{w}(\x)$}\\
    \,0 & \text{otherwise}
  \end{cases}
\vp
\end{equation}
While the simplicity of this idea makes it really enticing (and it has
in fact become quite popular), we anticipate that $\hat{w}(\x)$
is seldom accurate enough in practice, and this method generally
results in poor \emph{precision-recall} performance.

According to~\cite{Ventsel73} and the definition of $\RR$ in~\eqref{eq:ynor},
$P(\RR\mid\x,v)$ can be exactly written as: :
\begin{eqnarray}\label{eq:orprobs}
  P(\RR\mid \x,v)
   &=&\!\!\sum_{k=1}^n P(w_k=v\mid \x) \nonumber\\
   &-&\!\!\sum_{l<k} P(w_k=v, w_l=v\mid \x) \nonumber\\
   &+&\!\!\!\!\sum_{m<l<k}P(w_k=v, w_l=v, w_m=v\mid \x)\nonumber\\
   &\dots&\!\! (-1)^{n-1} P(w_1=v, \dots w_n=v\mid \x)
\vn
\end{eqnarray}

If the image regions are sufficiently small (e.g., line regions), it
can reasonably be expected that only one instance of each keyword may
appear in each region.  In these cases, all the joint probabilities in
Eq.\,\eqref{eq:orprobs} vanish and it simplifies to:
\begin{equation}\label{eq:sumApprox}
  P(\RR\mid \x,v) ~~\approx \sum_{1\leq k\leq n}\!P_{kv\x} 
\vn
\end{equation}
where, to simplify notation, $P_{kvx}$ is defined as:
\begin{equation}\label{eq:Pvkx}
  P_{kvx}~\eqdef~P(w_k=v\mid\x)
\end{equation}
Note that replacing the sum in Eq.\,\eqref{eq:sumApprox} with a %
\emph{probability average} would result in a deficient distribution
and therefore a worse approximation to $P(\RR\mid \x,v)$.
%
However, a drawback of Eq.\,\eqref{eq:sumApprox} is that, in the
(generally uncommon) cases where a keyword appears more than once in
an image region, $P(\RR\mid \x,v)$ may become improper since the sum
can be greater than one.

Therefore,
we propose a simpler approximation to Eq.\,\eqref{eq:orprobs} which,
as will be seen later, is better in general and, moreover, it is
intuitively appealing (see Fig.\,\eqref{fig:2Dpostgm}
and\,\eqref{fig:pgram} for illustration).
%
%
\begin{equation}\label{eq:maxApprox}
  P(\RR\mid \x,v) ~~\approx~ \max_{1\leq k\leq n}P_{kv\x} 
\end{equation}
A similar approximation has been used as a popular, good heuristic for
\emph{confidence estimation} in many works of automatic speech and
handwritten text recognition~\cite{Sanchis:11,ConfMeasuresHTR}, and
recently also for keyword indexing and search in~\cite{toselli16}.

A less simple, but hopefully better approximation to
Eq.\,\eqref{eq:orprobs} can be derived by using the naive Bayes
approximation to the joint probabilities of Eq.\,\eqref{eq:orprobs}:
\begin{equation}\label{eq:orprobsNB}
  P(\RR\mid \x,v)
   ~\approx\,\sum_{k=1}^n P_{kv\x}
          -\sum_{l<k} P_{kv\x}\,P_{lv\x}
      +\!\!\sum_{m<l<k}\!\!P_{kv\x}\,P_{lv\x}\,P_{mv\x}
      ~\dots~ (-1)^{n-1} P_{1v\x}\dots P_{nv\x}
\end{equation}
It can then be shown by simple induction that Eq.\eqref{eq:orprobsNB}
can be efficiently computed by dynamic programming according to the
following recurrence relation:
\begin{equation}\label{eq:recOrProbs}
P(\RR\mid \x,v) ~\approx~ q(n),\quad\text{where} ~~
q(k) = \begin{cases}\displaystyle
  \,P_{1v\x} & \text{if $\,k=1$}\\
  \displaystyle
  \,P_{kv\x} + q(k-1)(1 - P_{kv\x}) & \text{if $\,k>1$}\\
\end{cases}
\end{equation}
\subsection{Estimating Image-Region Relevance Probabilities
                  from Posteriorgrams}               \label{sec:YNfromPgram}

In proper KWS no transcript of $\x$ is available, but %
$P(w_k\!=\!v\mid \x)\equiv P_{kv\x}$ can be estimated from the
posteriorgram for $k\in\{1,2,\dots\}$.  To this end, we can divide the
whole region $\x$, into $n$ (maybe slightly overlapping or disjoint)
sub-regions or blocks, $B_1,\dots B_k,\dots B_n$, where a sufficiently
high and wide (usually rather flat) local maximum of \mbox{$P(v\mid
  \x,i,j)$} is observed in each $B_k$ for some $v\in V$ (see
Fig.\,\ref{fig:2Dpostgm}, where $n$ should be around 25 or, more
concretely, the uni-dimensional illustration of Fig.\,\ref{fig:pgram},
where $n$ would be 9 or 10).
%
Then,
\vn
\begin{equation}\label{eq:maxPkvx}\displaystyle
  P_{kv\x} ~~\approx~ \max_{(i,j)\in B_k}P(v\mid\x,i,j)
\vn
\end{equation}

This estimate can be used in any of the approximations
(\ref{eq:1bOrApprox},\ref{eq:sumApprox},\ref{eq:maxApprox},\ref{eq:recOrProbs})
of Sec.\,\ref{sec:regionYNprobs}.  But it becomes particularly simple
in~Eq.\,\eqref{eq:maxApprox}:
\begin{equation}\label{eq:ynFromPgram}
  P(\RR\mid \x,v) ~\approx
     \max_{1\leq k\leq n}\,\max_{(i,j)\in B_k} \!P(v\mid \x,i,j) ~=~
     \max_{i,j}P(v\mid \x,i,j)
\end{equation}
%
Note that the number of sub-regions, $n$, needed to derive
Eq.\,\eqref{eq:ynFromPgram}, finally becomes irrelevant.

Clearly, the maximization of Eq.\,\eqref{eq:ynFromPgram} can be
carried out during the process of computing \mbox{$P(v\mid \x,i,j)$}
itself and this approach does not add any computational cost to that
of obtaining the posteriorgram.
Moreover, this approach straightforwardly allows us to locate the
precise position of the spotted word within a spotted region $\x$,
just as the $\argmax$ of Eq.\,\eqref{eq:ynFromPgram}.  For instance,
if $\x$ is the image region of Fig.~\ref{fig:2Dpostgm}, spotted while
searching for the word \textbf{\texttt{"matter"}}, an instance of this
word can be readily located in a sub-region around $(i=150, j=110)$,
where $P(v\mid\x,i,j)$ is maximum.  Of course, the positions of the
other two instances of \textbf{\texttt{"matter"}} can also be easily
obtained from the posteriorgram.


Eq.\,\eqref{eq:ynFromPgram} heavily relies on the simplest
approximation to Eq\,\eqref{eq:orprobs}. 
As suggested in the above derivation of Eq\,\eqref{eq:ynFromPgram},
other more direct, posteriorgram-based approximations to
$P(\RR\mid\x,v)$ 
can be derived by actually dividing the image region $\x$, into $n$
sub-regions or blocks, $B_1,\dots B_k,\dots B_n$, by locating
adequately large and wide local maxima of \mbox{$P(v\mid \x,i,j)$}.
Then, all the other equations
(\ref{eq:1bOrApprox},\ref{eq:sumApprox},\ref{eq:recOrProbs}) of
Sec.\,\ref{sec:regionYNprobs} can be used to also efficiently and
perhaps more accurately compute \mbox{$P(\RR\mid \x,v)$}.
The relative empirical performance of all these approximations will be
reported in Sec.\,\ref{sec:results}. We anticipate, however, that our
best proposal will be Eq.\,\eqref{eq:ynFromPgram}, based on
~Eq.\,\eqref{eq:maxApprox}.
%

\begin{figure*}[htbp]
  \centering
  \hspace{-1em}%
  \includegraphics[height=.35\linewidth]{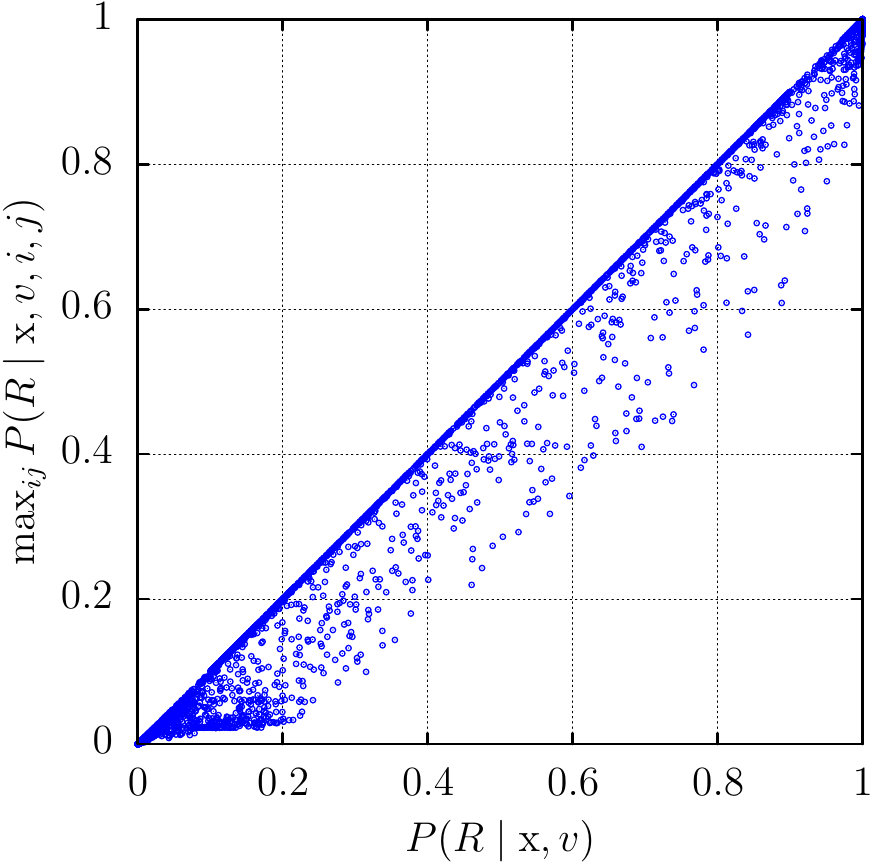}%
  ~~
  \includegraphics[height=.35\linewidth]{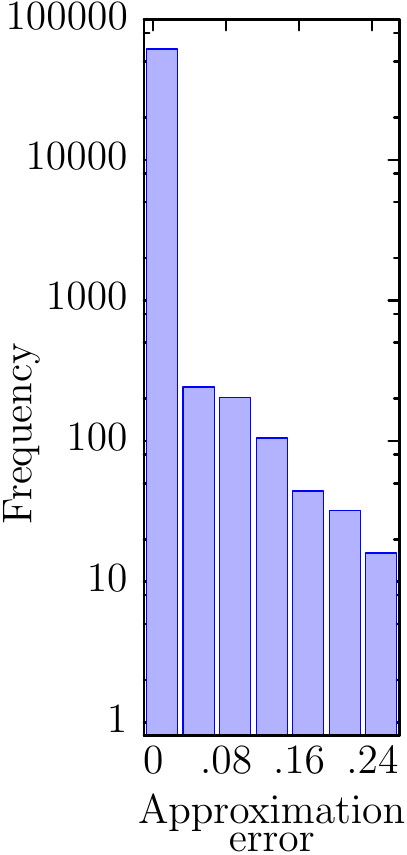}
\vspace{-0.5em}
  \caption{\small\label{fig:MxProbVsFrwProb} Correlation between exact
    and approximate relevance probabilities obtained for line regions
    and words of the Bentham test set.  $P(R\mid \x,v)$ is generally
    well approximated by $\max_{ij}P(R\mid\x,v,i,j)$ and the
    approximation error is almost null in more than 99.5\% of the
    regions and words spotted.}
\vspace{-0.5em}
\end{figure*}

Techniques similar to those proposed in Section~IV
of~\cite{toselli16b} can be used to compute (almost) exact relevance
probabilities (using Eq.\,\eqref{eq:relProbMain}, to be discussed in
Sec.\,\ref{sec:kwsHtr}).  In contrast to Eq.\,\eqref{eq:ynFromPgram},
this approach does not allow to obtain precise locations of the
spotted words within the spotted regions and, moreover, it is much
more complex and computationally demanding.
Nevertheless, for comparison purposes, we have afforded it on one of
the small test sets described in Sec.\,\ref{sec:datasets}.  The
results, summarized in Fig.\,\ref{fig:MxProbVsFrwProb}, show that
Eq.\,\eqref{eq:ynFromPgram} provides practically exact results for
more than 99.5\% of the (line) regions and words spotted in this
dataset.

From the above discussion, and from the results of
Sec.\,\ref{sec:results}, our concluding proposal will be to use
Eq.\,\eqref{eq:ynFromPgram}.  Not only it provides (almost) the best
approximation to the ideal relevance probabilities
(Eq.\,\eqref{eq:orprobs}), but it also leads to simple implementations
and, moreover, it straightforwardly allows to obtain accurate bounding
boxes for the spotted words within the spotted image regions.

\subsection{Line-Region Keyword indexing and Search}
                                                       \label{sec:lineKWS}

At the beginning of Section\,\ref{sec:regionKWS}, line image regions
were suggested as particularly adequate search targets for
handwritten word indexing and search.  More specifically, in connection
with the discussion in Sec.\,\ref{sec:pgram}, the following two main
advantages of these regions can be identified:
\vnnnn

\begin{enumerate}\itemsep=0em
\item \emph{Line regions} provide sufficiently \emph{rich}
  \emph{linguistic} \emph{context} to allow computing very precise
  word classification probabilities.
\item Line regions allow for very \emph{efficient computation} of
  posteriorgrams by smart \emph{choices} of the sets of relevant
  \emph{marginalization boxes}, $\mathcal{B}(i,j)$ and wise
  \emph{vertical} \emph{sub-sampling}.
\end{enumerate}

\vnn
For a line-shaped region, the relevant sets of \emph{marginalization}
\emph{boxes} needed to compute the posteriorgram according to
Eq.\,\eqref{eq:pgram0a} can be just defined by \emph{horizontal}
\emph{segmentation}.  As will be discussed later, very adequate sets
$\mathcal{B}(i,j)$ can be efficiently obtained as a byproduct of using
a holistic, context-aware handwritten recognizer on the whole line
image region.

On the other hand, in general, \emph{vertical sub-sampling} can just
reduce to guessing a proper line height and running a vertical-sliding
window of this height with some overlap.
However, in many cases, text line spacing is fairly regular and
standard line detection and extraction
techniques~\cite{Likforman:07,Li:08,Louloudis:09} can yield accurate
results.  This may lead to computation reductions and potentially
increase the precision.  The possible lack of robustness of this
approach can be easily alleviated by means of
\emph{over-segmentation}~\cite{Bluche:14,kumar2012simple}.
%

In what follows we will assume (as in~\cite{Kolcz:00,Terasawa:09,
Frinken12,Fischer12,Khayyat:14,Wshah:14,toselli16} -- see
also~\cite{Giotis:2017}) line-shaped image regions as our target
resolution level for keyword indexing and search.  Of course, once a
line spot is determined, the exact position(s) of the keyword searched
for within the line can be easily obtained as a byproduct (and/or through
simple post-processing).
%

\subsection{1-D Posteriorgrams and
                  Line-Region Relevance Probabilities} \label{sec:wgpost}



%
%

First, line images are assumed to be submitted to the usual line
preprocessing (and maybe feature extraction) steps adopted for
line-oriented
HTR~\cite{Bazzi99,Toselli04,Rodriguez09,Sanchez:16,Graves:09,Espana:11}.
This way, each text line image $\x$ becomes represented as a
sequence,
$x$, of vectors, where each vector $\vec{x}_i$, $1\leq i\leq m=|x|$,
describes grey-level values (or adequate features thereof) of a
narrow vertical box (or ``frame'') extracted at uniformly spaced
horizontal line positions.
The (line-level, 1-D) posteriorgram of $x$ is then (re-)defined as the
probability that $\vec{x}_i$ is one of the vectors representing a horizontal
segment of $\x$ which uniquely contains the word $v$.  It is denoted as:
\begin{equation}\label{eq:pgram}
  P(v\mid x, i),~~ 1\leq i\leq m,~ v\in V
\end{equation}


As in the general two-dimensional case, $P(v\mid x,i)$ can be easily
computed by considering that $v$ may appear in any horizontal segment
$\S$ 
of $x$ which includes the vector $\vec{x}_i$:
%
%
\begin{equation}\label{eq:pgram1}
  P(v\mid x,i)
  ~= \sum_{\S\in\mathcal{S}(i)}P(v,\S\mid x,i)
  ~= \sum_{\S\in\mathcal{S}(i)}P(\S\mid x,i)\,P(v\mid x,\S)
\end{equation}
where $\mathcal{S}(i)$ is the set of reasonably shifted and sized
segments which contain the frame $i$ and, as in
Eq.\,\eqref{eq:pgram0b}, $P(\S\mid x,i)$ can be assumed uniform and
replaced by an adequate constant.
%
%
Also in this one-dimensional setting, any system capable of
recognizing pre-segmented vector sequences corresponding to word
images should explicitly or implicitly rely on computing %
\mbox{$P(v\mid x,\S)$} and can thereby be used to obtain the
posteriorgram according to Eq.\,\eqref{eq:pgram1}.

\begin{figure}[htb]
\hspace{13.5mm}\resizebox!{6.6mm}{\includegraphics{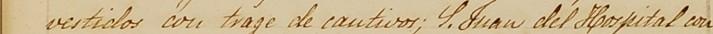}}
\vspace{-2mm}

\hspace{-7.0mm}\resizebox!{60mm}{\includegraphics{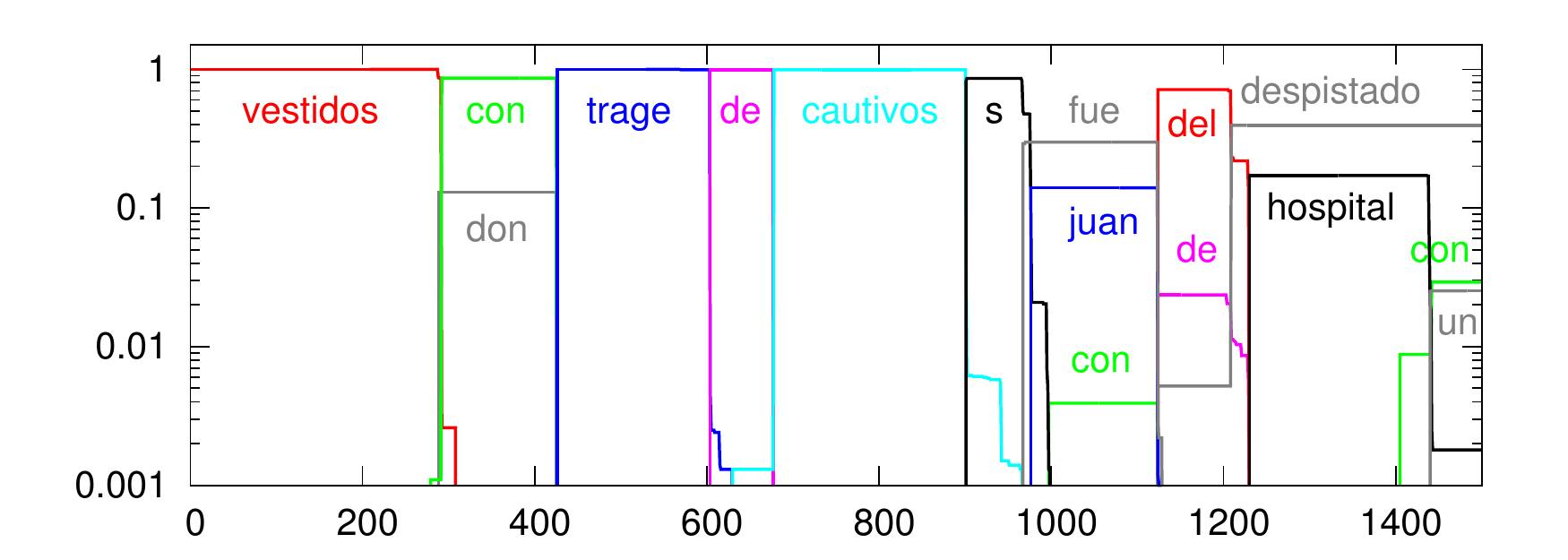}}
\vspace{-1em}
\caption{\small                                       \label{fig:pgram} %
  A 1-D posteriorgram obtained with the approach outlined in
  Sec.\,\ref{sec:approaches}, using a recognizer based on optical
  HMMs and a 2-Gram language model.}
\end{figure}

As discussed previously, an important advantage of line-level
processing is that it allows to easily take into account the rich
contextual word information provided by words surrounding each query
word.
%
%
In Sec.\,\ref{sec:approaches} we will discuss our concrete proposal to
efficiently obtain this kind of posteriorgrams using techniques
described in~\cite{toselli16}.
%
A real example of context-aware line image posteriorgram obtained in
this way is shown in Fig.~\ref{fig:pgram}.

\vp

%
%

Finally, Eqs.\,(\ref{eq:recOrProbs}--\ref{eq:ynFromPgram}) in
Sec.\,\ref{sec:regionYNprobs} and Sec.\,\ref{sec:YNfromPgram} can
straightforwardly rewritten to obtain line-region relevance
probabilities from $P(v\mid x,i)$.
In particular, Eq.\,\eqref{eq:ynFromPgram} becomes:
\begin{equation}\label{eq:ynFrom1Dpgram}
  P(\RR\mid x,v) ~\approx~ \max_iP(v\mid x,i)
\end{equation}

\subsection{Keyword Spotting and Handwritten Text Recognition}
\label{sec:kwsHtr}

Many authors in the field of KWS consider that KWS and HTR are very
different problems which should be tackled using distinct approaches
or methods.  Aiming to shed light on the relationship between these
two fields, in this subsection KWS will be re-visited from the HTR
point of view.  Following the probabilistic framework adopted
throughout this paper, we will see that both problems can be properly
and advantageously formulated according to identical probability
distributions.
This firmly suggests that well-established techniques to model and
train the probabilistic distributions used in the field of HTR can be
advantageously used for KWS, as well.
In fact, the most successful handwriting KWS methods (and those in
the closely related field of speech KWS, also known as Spoken Term
Detection), use in one way or another this relationship.

In Sec.\;\ref{sec:probFramework}, it was pointed out that KWS
essentially boils down to answering the question: ``is the word $v$
written in the text image region $\x$?''.
%
%
%
%
Clearly, a direct answer to this question is to check whether $v$
appears in a word sequence $w$ which constitutes the transcript of
$\x$.  But, since $w$ is unknown, it needs to be considered as the
value of a new random variable, $\WW$, defined over all the possible
transcripts of $\x$.
This allows us to obtain the KWS relevance probability by
marginalization on $\WW$:
%
%
\begin{eqnarray}\label{eq:transcrMarginal}
P(\RR\mid \x,v)
 ~=~ \!\!\sum_w 
             P(\RR,\WW=w \mid \x,v)
 ~=~ \!\!\sum_w 
             P(\RR\mid w,\x,v) \cdot P(w \mid \x,v)
\end{eqnarray}
where $\WW$ has been omitted as previously done with other random
variables and $w$ ranges over the set of all sequences of words in $V$.
%
%
Now, since $w$ is given in $P(\RR\mid w,\x,v)$, it can be written as:
%
\begin{equation}\label{eq:relCondTranscr}
P(\RR\mid w,\x,v) ~=~
\begin{cases}
    1 & v\in w\\
    0 & \text{otherwise}
  \end{cases}
\vspace{0.3em}
\end{equation}
%
On the other hand, since image transcripts ($w$) are independent form
user queries ($v$), the term $P(w\mid\x,v)$ in
Eq.\,\eqref{eq:transcrMarginal} simplifies to $P(w\mid\x)$ and we can
write:
%
%
%
\begin{equation}\label{eq:relProbMain}
P(\RR\mid\x,v) ~=\!\sum_{w:v\in w} P(w\mid\x)
\end{equation}
That is, KWS relevance probabilities can be properly computed on the
base of the probability that a sequence of words $w$ is the transcript
of the image region $\x$.
Interestingly, this is exactly the same distribution used by modern
Handwritten Text Recognition systems, which search for a most likely
transcript of the given text image region $\x$ according to the minimum
Bayes error criterion~\cite{Duda73}; that is:
%
%
\begin{equation}\label{eq:htrMain}
\hat{w} ~=~ \argmax_w P(w\mid\x) 
\end{equation}
%


Directly computing KWS relevance probabilities according to
Eq.\,\eqref{eq:relProbMain} constitutes a complex computational
problem.  It can be solved by means of a dynamic programming technique
similar to the ``forward'' approach proposed in Section~IV
of~\cite{toselli16b}.  But, even using this technique, the
computational cost is still very high, as the results presented
in~\cite{toselli16b} clearly show.  Therefore, in this paper, we will
omit these computational details and will only use this approach for
comparison purposes.

Nevertheless, adequate approximations can be obtained by extending the
naive ``1-best'' idea based on Eq.\,\eqref{eq:htrMain}, discussed in
Sec.\,\ref{sec:regionYNprobs}.
More specifically, the sum in Eq.\,\eqref{eq:relProbMain} is
restricted to a set of $n$ word sequences for which $P(w\mid\x)$ is
largest.  This set, often called ``$n$-best list'', can be obtained as
a byproduct of solving the optimization problem of
Eq.\,\eqref{eq:htrMain}~\cite{Jelinek98}.  If $n$ is large enough,
accurate relevance probabilities can be obtained, but computational
costs grow exceedingly fast for increasing values of $n$.  While using
$n$-best lists for KWS is promising, more research work is required
and we will not elaborate further in this specific direction in the
present paper.
Instead, as discussed in the previous section, we stick to
posteriorgrams efficiently and accurately obtained using word-graph
representations of the distribution $P(w\mid\x)$~\cite{toselli16}.

\vppp
\section{Proposed Statistical Framework and
            KWS Approaches}                \label{sec:approaches}

In the framework proposed in Sec.\,\ref{sec:pgram}, a KWS method is
assumed to implicitly or explicitly visit all the image locations
$(i,j)$ and all the possible bounding boxes (BB) $\B$ containing
$(i,j)$ (or adequately selected locations and/or BBs thereof).  For
each $(i,j)$ and $\B$, an (isolated) word recognizer or word-matching
technique of some kind is used to estimate the posterior probability,
$P(v\mid\X,\B)$, that a given keyword $v$ is the (only) word written
in $\B$.  Then Eq.\,\eqref{eq:pgram0b} is somehow used to compute the
pixel-level posteriorgrams, from which region relevance probabilities
are computed as explained in Sec.\,\ref{sec:regionYNprobs} and
Sec.\,\ref{sec:YNfromPgram}.

A single $\B$ encompasses $O(I\,J)$ pixels.  Thus, estimating
$P(v\mid\X,\B)$ for all $v\in V$ can be assumed to be at least
$\Omega(N\,I\,J)$, where $N$ is the number of keywords.  In general,
for each image location there are $O((IJ)^2)$ possible $b$'s, and the
number of locations is $O(I\,J)$.
Therefore, the overall computational complexity of \emph{directly}
computing a posteriorgram in this way, for all the $I\!\cdot\!J$
pixels in a full image, is really huge: at least $\Omega(N\,(IJ)^4)$.

It is then no surprise that the history of development of KWS for text
images can be interpreted in terms of how to deal with the different
aspects of this exorbitant computational cost.

\subsection{Interpretation of Other KWS Methods}

According to the framework introduced in this paper, four main aspects
can be identified which characterize most (QbS) KWS methods for
(handwritten) text images proposed so far.

\vn
\begin{itemize}\itemsep=-0.3em
\item How to effectively sample the exceedingly large number of pixel
  locations of $\X$
\item How to adequately define the set $\mathcal{B}(i,j)$ of
  marginalization BBs required in Eq.\,\eqref{eq:pgram0b}
\item How to deal with the summation in Eq.\,\eqref{eq:pgram0b}
\item How to estimate the word classification posterior $P(v\mid\X,\B)$
      for each $\B\in\mathcal{B}(i,j)$
\vnn
\end{itemize}


We start discussing the first three aspects, which are closely
inter-related and together aim to deal with the computational costs
which essentially depend on the image size, $I\!\cdot\!J$.

All of the early KWS techniques relying on pre-segmented word
images~\cite{Giotis:2017} circumvented the prohibitive cost of
computing Eq.\,\eqref{eq:pgram0b} by reducing the summation to just
one fixed word-sized image region or ``patch''.  Moreover, KWS
``scores'' (proxy for word posteriors) are computed only at the
relatively small number, $l$, of previously given locations of these
word patches.  This provides a simplistic solution to all the first
three aspects discussed above.
Obviously, by naively assuming perfect word image detection and
segmentation, the computational cost is dramatically reduced down to
$O(N\,l)$, which clearly explains the mighty popularity of this
simplistic idea.

More recent works, such as~\cite{Rodriguez-Serrano:09}, rely on
automatic over-segmentation of the text images to mitigate the impact
of word segmentation errors.  Such techniques rely on a richer, more
realistic sub-sampling and, to some extent, go towards approximating
the marginalization in Eq.\,\eqref{eq:pgram0a} and~\eqref{eq:pgram0b}.


In full segmentation-free KWS methods~\cite{Giotis:2017}, sub-sampling
is generally performed through a \emph{sliding-window} sweep over the
image -- see, e.g.,~\cite{Hast:16}.  However, full pixel-by-pixel
sweep is again much too expensive and, in many works, an adequately
small number $p$ of 
\emph{key-points} which define possible parts of the objects of interest
(words), are previously located~\cite{Hast:16,Giotis:2017}.  This way,
assuming marginalization is simplistically reduced to just one
candidate BB or ``patch'' -- which is usually the case, computational
cost can be reduced down to $O(N\,p)$ where, in general, $p\!>\!\!>l$.

On the other hand, in KWS approaches which work with (unsegmented)
\emph{line} image regions, $\x$, the summation in
Eq.\,\eqref{eq:pgram0b} becomes uni-dimensional (i.e.,
Eq.\,\eqref{eq:pgram1}).  In many of these approaches the sum in this
equation is more or less explicitly approximated only by the
dominating addend (which is typically a good approximation --
generally much better than relying on a single, given BB).  Then,
\emph{dynamic programming} techniques are used to avoid repeated
computations during a sliding window process over the horizontal
positions of $\x$.  This is specifically the case of
\mbox{(word-)seg}\-mentation-free \emph{dynamic time warping} KWS
methods such as~\cite{Terasawa:09,Kolcz:00}, as well as all the modern
techniques based on HMMs~\cite{Thomas:10,Fischer12,Wshah:12} and
recurrent neural networks~\cite{Frinken12}.

Nevertheless, obtaining a true full 1-D posteriorgram for each of the
$L$ line-regions in $\X$ would still entail a formidable amount of
computation.  In general, the size of of the set of marginalization
segments, ${S}(i)$, is $O(m^2)$ and the the average segment length is
$\Theta(m)$, where $m$ is the length of $x$, which in turn is $O(I)$.
Therefore, even if repeated computations are avoided, the overall
asymptotic time complexity is $O(NLI^2)$.
%
Fortunately, in this simpler 1-D case, reasonably good and
computationally cheaper approximations can be obtained in a variety of
ways.  Below we will outline the approach we propose and we have used
to obtain the results reported in Sec.\,\ref{sec:results}.

Let us now discuss the last aspect which characterize a KWS; namely how
to estimate the word classification posteriors $P(v\mid\X,\B)$.  Three
main approaches can be identified: HMMs, (Recurrent) Neural Networks
(RNN) and distance-based.


The most common use of HMMs is to model a word $v$ as an explicit or
implicit concatenation of character HMMs.  The combined HMM estimates
the likelihood that a word image patch $(\X,\B)$ is a rendering of the
modeled word; i.e., $P(\X,\B\mid v)$, which is proportional to
$P(v\mid\X,\B)$ assuming $P(v)$ and $P(\B)$ are uniform.
Many popular approaches, such as the \emph{``filler''} or
\emph{``garbage''} models~\cite{Fischer12}, fall into this category.

Let us now focus on RNN~\cite{Frinken12}.  For a given word image
patch $(\X,\B)$, these networks directly provide a sequence of
posterior probabilities $P(c\mid\X,\B,i)$ where $c$ is a character and
$i$ is a horizontal position within $\B$. For a keyword $v$, composed
of characters $c_1,\dots,c_K$, dynamic programming can be used to
obtain a best matching path $\Phi(\cdot)$, which assign each position
$i$ to one of the $K$ characters of $v$.  Then usual independence
assumptions lead to the naive Bayes approximation:
$P(v\mid\X,\B)\approx\prod_{i}P(c_{\Phi(i)}\mid\X,\B,i)$, where $i$
ranges over the horizontal positions of $\B$.

Finally, many early approaches to KWS, notably segmentation-based
ones, are \emph{based on distances} between representations of queries
and images.  It is well known that distances can be used to
approximate probability distributions in several ways~\cite{Duda73}.
If $y$ and $z$ are, respectively, representations of a query word $v$
and an image BB $(\X,\B)$, then a very simple estimator of the
classification posterior required in~Eq.\,\eqref{eq:pgram0b} is: %
$P(v\mid\X,\B)\approx\phi(y,z)/\sum_u\phi(u,z)$, where
$\phi(u,z)=\exp(-d(u,z))$, $\,d(\cdot,\cdot)$ is the distance, and $u$
ranges over (an adequate set of) query word representations.

Distance-based KWS methods~\cite{Giotis:2017} often drop the
denominator and use just unnormalized ``scores''.  While the resulting
lack of basic probabilistic properties may not change the
\emph{individual} average precision of each query (see
Sec.\,\ref{sec:evalMeas}), unnormalized scores are prone to more or
less severely hinder the global average precision for a %
\emph{set of queries}.  This problem may also affect HMM and RNN
methods based on heuristically defined relevance probabilities, rather
than using a proper approximation to $P(\RR\mid\x, v)$.  This leads
some authors (e.g.,~\cite{Fischer12}) to resort to using %
\emph{``local thresholds''} in their evaluation protocols.

Most distance-based methods are QbE, for which many representation
schemes and metrics have been proposed~\cite{Giotis:2017}, but some
recent QbS proposals such as~\cite{almazan14} are also distance-based.




\subsection{Proposed KWS Approach}

To finish this section, we discuss here the specific approach we
propose to compute accurate, context-aware line-region posteriorgrams,
and the corresponding image region relevance probabilities, in an
effective and efficient way.  It formally follows the statistical
framework developed in previous sections and, as previously mentioned,
is based on techniques introduced in~\cite{toselli16}.
The main idea is to use a \emph{Word Lattice} or \emph{Graph}
(WG)~\cite{Ort97,toselli16}, obtained as a byproduct of solving
Eq.\,\eqref{eq:htrMain} (see Sec.\,\ref{sec:kwsHtr}) by means of an
adequate 
handwritten text recognizer~\cite{Romero12a,toselli16}.
A WG of a (line) image-region, $\x$, is a very compact representation
of huge amounts of alternative image transcription results, including
the probability of each of the (millions of) hypothesized words and
the corresponding word segmentation boundaries.

The posteriorgram $P(v\!\mid\! x,i)$ can be obtained from a WG of $\x$
following essentially the same arguments as in
Eq.\,\eqref{eq:sumApprox} and Eq.\,\eqref{eq:pgram1}.
%
The basic concept is to consider that, for each position $i$, the
``relevant, reasonably shifted and sized'' segments in
$\mathcal{S}(i)$ are those given by the word segmentation hypotheses
associated with all the WG edges, $e$, labeled with the word $v$ and
such that $i$ is included within the segmentation boundaries specified
by departing and ending nodes of $e$.
See~\cite{toselli16} for more details about this process.
These word boundaries are
generally very accurate, not only for the words in the best hypothesis
of the WG (called the ``1-best'' transcript), but also for most of the
edges associated with high-probability paths of the WG.  Therefore these
boundaries and probabilities provide highly informative data to allow
very accurately computing Eq.\,\eqref{eq:pgram1}.

For a line image-region of length $I$, the computational cost of
obtaining a posteriorgram 
in this way is in $\Theta(\kappa\,I)$, where $\kappa$ is a constant
which depends on the size of the WG~\cite{toselli16}.  According
to~\cite{toselli16,toselli17a}, this cost is generally negligible as
compared with the cost of producing the WG itself when using an
$N$-gram based, large vocabulary handwritten text recognizer.


The four aspects which characterize a KWS method are now briefly
outlined for the proposed approach: To cope with the exceedingly large
number of pixel locations in $\X$, first the image is sampled
vertically by adopting line image regions as discussed in
Sec.\,\ref{sec:lineKWS}, and horizontally according to the
segmentation boundaries included in the line image region WGs.
Similarly, we let the word segments represented in each WG define the
sets of marginalization BBs.  With these data, we compute the sum of
Eq.\,\eqref{eq:pgram1} (or Eq.\,\eqref{eq:pgram0b}) in
$\Theta(\kappa\,I)$ as explained in~\cite{toselli16}.  Finally, as
commented before, the word classification posteriors $P(v\mid\X,\B)$
are directly obtained from the WG edges~\cite{toselli16}
(probabilities which, in turn, were computed essentially as discussed
above for HMMs or RNNs).

Given a posteriorgram obtained in this way, the relevance
probabilities needed for keyword indexing are obtained according to
Eq.\,\eqref{eq:ynFrom1Dpgram}, or any of the 1-dimensional versions of
Eqs.\,(\ref{eq:recOrProbs}--\ref{eq:ynFromPgram}).  So, once the WGs
of the $L$ extracted line image regions are available, the overall
computational effort per page image is $O(N\,\kappa\,L\,I)$.  More
details about these costs, including those of WG generations can be
seen in~\cite{toselli16,toselli17a}.
%


\section{Experimental Framework}          \label{sec:experiments}

To evaluate empirically the KWS performance for the different proposed
approaches, in this and the following section will be described the
evaluation measures, benchmark datasets, query sets and experimental
setup for both RNN and HTR optical modeling and different methods for
computing the query relevance probability.

\subsection{Evaluation Measures}         \label{sec:evalMeas}

Let $\mathcal{Q}$ be a set of (word) queries and $\tau$ be a relevance
threshold.  The recall, $\rho(q,\tau)$, and the raw (non-interpolated)
precision, $\pi'(q,\tau)$, for a given query $q\in\mathcal{Q}$ are
defined as:
\begin{equation}
  \rho(q,\tau) ~=~ \dfrac{h(q,\tau)}{r(q)}\,,
  \qquad\quad
  \pi(q,\tau)  ~=~ \dfrac{h(q,\tau)}{d(q,\tau)}
  \label{eq:rho-pi-local}
\end{equation}
where $r(q)$ is the number of test image regions which are
\emph{relevant} for $q$ (according to the ground truth), $d(q,\tau)$
is the number of regions retrieved or \emph{detected} by the
system with relevance threshold $\tau$ and $h(q,\tau)$ is the number
of detected regions which are actually relevant (also called
\emph{``hits''}).

The interrelated trade-off between recall and precision
can be conveniently displayed as the so-called \emph{recall-precision}
(R-P) curves, $\pi_q(\rho)$~\cite{Egghe:08}.
Any KWS system should allow users to (more or less explicitly)
regulate $\tau$ in order to choose the R-P operating point which is
most appropriate in each query.  Good systems should achieve both high
precision and high recall for a wide range of values of $\tau$.
%
A commonly accepted scalar measure which meets this intuition is the
area under the R-P curve, $\pi_q(\rho)$, here denoted as $\bar{\pi}_q$
and called (raw) \emph{average precision}~\cite{Zhu:04,Robertson:08}.
%
In addition, to consider all the queries in $\mathcal{Q}$, the (raw)
\emph{mean average precision} (mAP, denoted as $\bar{\bar{\pi}}$) is
used:
\begin{equation} \label{eq:mAP}
 ~~~~~~~~~~~~~~
 \bar{\pi}_q ~= \int_0^1\!\!\pi_q(\rho)d\rho\,,
 ~~~~~~
 \bar{\bar{\pi}}~=~\frac{1}{|\mathcal{Q}|}\sum_{q\in\mathcal{Q}}\bar{\pi}_q
 ~~~~~~~~~~~~~~~~~~~~~~ \text{(mAP)}
 \!\!\!\!\!\!\!\!\!\!\!\!\!\!\!\!\!\!\!\!\!\!\!\!
\end{equation}
Obviously,
the mAP is undefined if $\exists q\in\mathcal{Q}$ for which
$\bar{\pi}_q$ is undefined, which happens if $r(q)=0$, that is, if no
test-set image region is relevant for $q$.
%
On the other hand, Eq.\,\eqref{eq:mAP} equally weights all the
queries, thereby ignoring the different amounts of relevant regions
for different queries.

To circumvent both of these issues, a global averaging scheme can be
adopted by computing the total number of test image regions which are
\emph{relevant} for all $q\in\mathcal{Q}$, the total number of regions
\emph{detected} with relevance threshold $\tau$ and the total number
of \emph{hits}, respectively, as:
\begin{equation}
  r =\!       \sum_{q\in\mathcal{Q}}r(q)\,,
  ~~~~~
  d(\tau) =\! \sum_{q\in\mathcal{Q}} d(q,\tau)\,,
  ~~~~~
  h(\tau) =\! \sum_{q\in\mathcal{Q}} h(q,\tau)
  \label{eq:rho-pi-global}
\end{equation}
Then the \emph{overall} recall and raw precision, and the (often
preferred) \emph{global average precision}, $\bar{\pi}$, referred to
as AP, are defined as:
\begin{equation}
  \rho(\tau) ~=~ \dfrac{h(\tau)}{r}\,,
  ~~~~~~~
  \pi(\tau)  ~=~ \dfrac{h(\tau)}{d(\tau)}\,,
  ~~~~~~~
  \bar{\pi} ~= \int_0^1\!\!\pi(\rho)d\rho
  ~~~~~~~~~~~~~~~~~ \text{(AP)}\!\!\!\!\!\!\!\!\!\!\!\!\!\!\!\!
\label{eq:AP}
\end{equation}

Even with this averaging scheme, raw precision can still be
ill-defined in some extreme cases and, moreover, raw R-P curves can
present an undesired distinctive saw-tooth shape~\cite{Egghe:08}.
Both of these issues are avoided by the so-called %
\emph{interpolated precision}, defined as:
%
\begin{equation}
  \pi'(\rho) ~=~ \max_{\rho':\rho'\geq\rho}\pi(\rho')
  \label{eq:interpol-precision}
\end{equation}
Intuitive arguments in favor of $\pi'(\rho)$, which is often adopted
in the literature, are discussed in~\cite{Manning:2008}.

The same interpolation scheme can be applied to a single query $q$,
resulting in the interpolated R-P curve $\pi'_q(\rho)$.  Then, the
\emph{interpolated} versions of mAP and AP are straightforwardly
computed using $\pi'_q(\rho)$ and $\pi'(\rho)$, rather than
$\pi_q(\rho)$ and $\pi(\rho)$, in Eq.\,\eqref{eq:mAP} and
Eq.\,\eqref{eq:AP}, respectively.

The use of interpolated precision becomes even more necessary for fair
evaluation of KWS results of the naive $1$-best KWS approach (see
Sec.\,\ref{sec:regionYNprobs}, Eq.\,\eqref{eq:1bOrApprox} and
Sec.\,\ref{sec:results}), where relevance probabilities are $1$ or
$0$, independent of $\tau$.  Therefore, in the raw R-P curve, only one
R-P point, $(\rho_0,\pi_0)$, is defined and the resulting raw AP would
be $0$, thereby preventing comparison with other KWS approaches.
In contrast, the interpolated precision curve becomes
$\pi'(\rho)\!=\!\pi_0$ \mbox{if~$0\!\leq\!\rho\!\leq\!\rho_0$,}~
\mbox{$\pi'(\rho)\!=0\,$} otherwise, with a resulting interpolated AP:
$\bar{\pi}=\pi_0\cdot\rho_0$.
\subsection{Datasets}                               \label{sec:datasets}

The main experiments were conducted on two large handwriting
datasets, called PLANTAS and BENTHAM.  In addition comparative
experiments were carried out on smaller, more usual benchmarking
datasets; namely, IAM, George Washington and PARZIVAL.  In all the
cases, ground-truth line segmentation is available, both for training
and testing images, and it is used in the experiments.  We leave for
future works to experiment with the vertical sub-sampling and
over-segmentation ideas discussed in Sec.\,\ref{sec:lineKWS}.  The
main features of the datasets used in the experiments are described in
the following subsections.

\subsubsection*{PLANTAS Dataset}

\begin{figure*}[htbp]
  \centering
  \includegraphics[width=.22\linewidth]{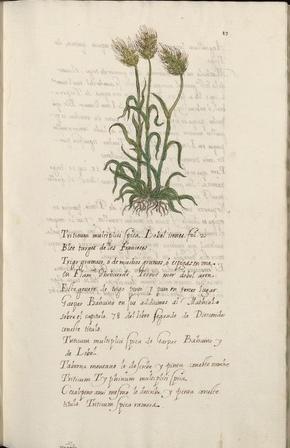}
  \hspace{.2em}%
  \includegraphics[width=.22\linewidth]{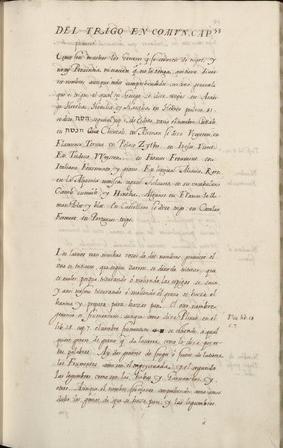}
\vspace{-0.5em}
  \caption{Examples of pages from the first volume of PLANTAS.}
  \label{fig:eje-Plantas}
\vspace{-0.5em}
\end{figure*}

The seven volumes of the book ``Historia de las plantas'' (hereinafter
shortened as PLANTAS) were written using a quill-pen by Bernardo de
Cienfuegos, one of the most outstanding Spanish botanists in the XVII
century.
The book was writing mainly in Spanish, but a significant number of
words and full sentences are in Latin and many other languages.
Today, this manuscript has become a source of valuable
information for research, specially for those interested in the
botanical knowledge of that historical period. The originals of
PLANTAS are currently available at the %
\emph{Biblioteca Nacional de España}, and a digital reproduction of it
can be found at the Biblioteca Digital Hisp\'anica%
\footnote{\url{http://bdh-rd.bne.es/viewer.vm?id=0000140162}}. Examples
of page images of this dataset are shown in
Fig.\,\ref{fig:eje-Plantas}.
In this work, only the first volume of PLANTAS (Mss\,3357, with
$1\,035$ pages and around $20\,000$ handwritten text lines) was
considered for experimentation.
%
%
%

Using techniques described in~\cite{toselli10a,Romero12a}, PLANTAS
was computer-assisted transcribed into tagged text which concomitantly
captured both the \emph{diplomatic} and \emph{``modernized''}
transcripts, and also certain \emph{semantic} clues~\cite{toselli17}.
The diplomatic facet captures, as accurately as possible,
all what can be seen in the text image (abbreviations, underlined and
crossed-out text, interline inserted text, etc.).  The modernized
component provides abbreviation expansions, normalized capitalization
and punctuation, etc.  The semantic tags, finally, describe the
language of each non-Spanish word and some word functions such as
catch-words and hyphenation.
%
%
In order to improve the computer-assisted experience and effectiveness,
tagged transcripts were used as such (tags and dual text included) to
increasingly train the language models used for interactive
transcription~\cite{toselli17}.  As discussed later on, the language
model used for the experiments presented in this paper was also
trained with the fully tagged transcripts of the training images.
But only tag-free, \emph{diplomatic} transcripts were employed to
obtain the KWS results reported in this paper.  See details in
Sec.\,\ref{sec:expSetup}.

Table\,\ref{table:PLANTAS_stat} shows experimental partitions and
general statistics of the PLANTAS tagged dataset. Glossaries, indices,
blank page images and images containing only drawings are ignored in
the ``Image'' data.
The rows ``Running tokens'' and ``Running OOV'' show the total number
of tagged words and the percentage of Out-Of-Vocabulary (OOV) tokens,
respectively.
The OOV tokens in the Validation column correspond to tokens which do
not appear as identically tagged words in the training set, while
those in the Test column are tokens that do not appear in the training
or in the validation sets.
Rows labeled ``Text'' contain data of the training partition which
additionally includes extra text, aimed at enhancing language model
training, extracted from the glossaries and indices of the excluded
page images.
Rows labeled ``Diplomatic'' show the corresponding partitions and
statistics of diplomatic reference transcripts used in the KWS
experiments reported in this paper.  The running words (and lexicon)
include the extra diplomatic text from glossaries and indices.
Note that, in contrast with the other datasets discussed below, in this
part of Table\,\ref{table:PLANTAS_stat} the text is assumed to be
\emph{tokenized}; that is punctuation marks are separated and
considered as ``words''.
In addition, validation and test \emph{word error rate} (WER),
obtained by a RNN-based HTR system using the same optical and language
models as for KWS (see Sec.\,\ref{sec:expSetup}), is reported.

The relative sizes of partitions were decided taking into account the
experience acquired during the assisted transcription process which
was followed to create the dataset ground-truth.
More details about this process and the resulting dataset appear
in~\cite{toselli17}.


\begin{table}[htbp]
  \centering
  \caption{
    \label{table:PLANTAS_stat}%
    The PLANTAS tagged dataset. Rows labeled ``Text'' correspond
    to the training image transcripts plus extra text extracted
    from glossaries and indices, used to train the language model.
  }
  \vspace{0.7em}
  \setlength{\tabcolsep}{3pt}
  \begin{tabular}{cl|rrr|r}
    \toprule[1.5pt]
    & Number of:         &~~Training& Validation & Test  & Total    \\
    \midrule[1.2pt]
    \multirow{5}{*}{\rotatebox{90}{Image}}
    & Pages              &      224 &      40 &      607 &      871 \\
    & Lines              &   6\,788 &     955 &  11\,801 &  19\,544 \\
    & Running tokens     &  67\,912 &  9\,753 & 117\,029 & 194\,694 \\
    & Token set size     &  10\,861 &  2\,198 &  14\,018 &  20\,834 \\
    & Character set size &       76 &      76 &       76 &       76 \\
    \midrule[1.2pt]
    \multirow{3}{*}{\rotatebox{90}{Text}}
    & Running tokens     &  70\,201 &  9\,753 & 117\,029 & 196\,983 \\
    & Token set size     &  11\,890 &  2\,198 &  14\,018 &  21\,417 \\
    & Running OOV\,(\%)  &       -- &     8.95 &   11.68 &       -- \\
    \midrule[1.2pt]
    \multirow{4}{*}{\rotatebox{90}{Diplomatic~}}
    & Running words      & 70\,201  & 9\,753 & 117\,026 & 196\,980 \\
    & Lexicon size       & 11\,045  & 2\.126 &  13\,124 &  19\,693 \\
    & Running OOV\,(\%)  &       -- &   8.40 &    10.81 &       -- \\
    \cmidrule[0.1pt]{2-6}
    & WER\,(\%)          &       -- &   15.62 &    19.61 &       -- \\
  \bottomrule[1.5pt]
  \end{tabular}
\end{table}



\subsubsection*{BENTHAM Dataset}
\vspace{-0.5em}

The whole set contains more than $80\,000$ images of manu\-scripts
written by the renowned English philosopher and reformer Jeremy
Bentham (1748-1832) and his secretarial staff~\cite{Causer:12b}.
It includes texts about legal reform, punishment, the constitution,
religion, and his famous \emph{``panopticon''} prison paradigm.
%

%

\vspace{-0.5em}
\begin{figure}[htbp]
  \centering
  \includegraphics[height=.35\linewidth]{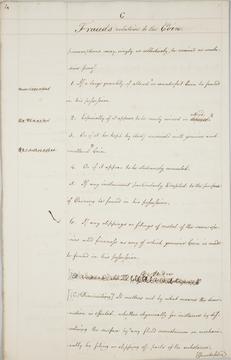}%
  \hspace{1em}
  \includegraphics[height=.35\linewidth]{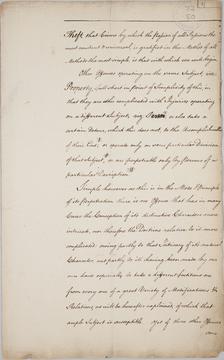}
  \vspace{-0.5em}
  \caption{Examples of BENTHAM page images.}\label{sample}
\vspace{-0.5em}
\end{figure}

From the Bentham data currently available, only a relatively small and
simple set of $433$ page images, written by an unknown number of
different writers,
is used in this work (see examples in Fig.~\ref{sample}). It is
exactly the same dataset used in the ICFHR-2014 HTRtS competition%
\footnote{%
\url{www.transcriptorium.eu/~htrcontest/contestICFHR2014/public_html/HTRtS2014}
} %
of handwritten text recognition~\cite{Sanchez:14a}.
These pages contain $11\,473$ lines with nearly $110\,000$ running
words and a vocabulary of more than $9\,500$ different words. The last
column in Table~\ref{table:BENTHAM_firstBatch} summarizes the basic
statistics of these pages.  It should be noted that these statistics
correspond to \emph{not tokenized} text; that is, each non-blank
sequence of characters is assumed to be a ``word'', even though it may
often correspond to a real word plus a punctuation mark.
The dataset was divided into three subsets for training, validation
and test.
Unlike PLANTAS, which is a single-writer manuscript, BENTHAM was
written by many hands, with a large variability in their handwriting
styles. Therefore it requires much more training samples to adequately
train writer-independent character models.
%
%
%
Table~\ref{table:BENTHAM_firstBatch} shows details of these partitions.

%
\begin{table}[htbp]
  \vspace{-1em}
  \centering
  \caption{
    \label{table:BENTHAM_firstBatch}%
    The Bentham dataset, used in the ICFHR-2014 HTRtS
    competition~\cite{Sanchez:14a}.
  }
  \vspace{2mm}
  \setlength{\tabcolsep}{3pt}
  \begin{tabular}{l|rrr|r}\toprule[0.1em]
    Number of:         & Training & Validation & Test &  Total   \\
    \midrule
    Pages              &      350 &      50 &      33 &     433  \\
    Lines              &   9\,198 &  1\,415 &     860 &  11\,473 \\
    Running words      &  86\,075 & 12\,962 &  7\,868 & 106\,905 \\
    Lexicon size       &   8\,658 &  2\,709 &  1\,946 &   9\,716 \\
    Character set size &       86 &      86 &      86 &       86 \\
    Running OOV(\%)    &     --   &     6.61 &    5.30 &    --    \\
    \midrule
    WER(\%)            &       -- &   13.17 &   19.76 &       -- \\
    \bottomrule[0.1em]
  \end{tabular}
\end{table}

Finally, validation and test WER, obtained by a RNN-based HTR system
using the same optical and language models as for KWS (see
Sec.\,\ref{sec:expSetup}), are also reported in this table. For more
details about this dataset refer to~\cite{Sanchez:14a}.

%



\subsubsection*{IAM, George Washington and Parzival datasets}

IAM is a publicly available, well known modern English handwritten
text corpus, compiled by the FKI-IAM Research Group on the base of the
Lancaster-Oslo/Bergen Corpus (LOB).  The last released version (3.0)
is composed of $1\,539$ scanned text pages, handwritten by $657$
different writers and partitioned into writer-independent training,
validation and test sets. The line segmentation provided with the
corpus~\cite{Marti:02} is used here.
Statistics of the IAM dataset appear in
Table~\ref{tab:corpora-iamdb}. In this case for this dataset, the
``text'' information refers to three external text corpora (LOB,
Brown, and Wellington, collectively called ``LBW'') which were
employed for compiling the lexicon and for training the IAM 
language model~\cite{Bertolami:08}.

\textsc{Parzival} (PAR) is a database containing $45$ digital images
of a medieval manuscript from the 13th century written down in Middle
High German language~\cite{Fischer:09}.
Although several writers have contributed to the manuscript, all the
writing styles found in the data set are very similar.
Table\,\ref{tab:corpora-others} presents also information about the
employed partition definition and statistics of this data set.

\textsc{George Washington} (GW) is a database which includes $20$
pages of letters~\cite{Lavrenko:04} written by George Washington and
his associates in the year $1\,755$. These $20$ relatively clean
pages, which exhibit a very similar writing style, have been selected
from a larger collection of images.
%
Because of the small size of the data set, four-fold cross validation
has been adopted for experimental evaluation.  Therefore,
Table\,\ref{tab:corpora-others} includes rounded mean values over all
cross validation sets.

Following the standard presentation of these benchmark corpora, as in
the case of BENTHAM, all the data in Tables\,\ref{tab:corpora-iamdb}
and~\ref{tab:corpora-others}, correspond to \emph{not tokenized} text.

\begin{table}[htbp]
\vspace{-0.5em}
  \centering
  \caption{
    \label{tab:corpora-iamdb}%
    The IAM dataset and the corresponding partition.
    The IAM Running words, Lexicon and OOV (out of vocabulary)
    figures labeled ``Text'' correspond to the external text
    corpora (LBW) used to train the language model.
  }
  \vspace{0.5em}                                       
  \tabcolsep=6pt
  {
    \begin{tabular}{l|l|rrrr}\toprule[0.1em]
      & Number of: & Training& \!Validation\!\!\! &~~~~~Test&Total \\
      \midrule
      \multirow{4}{*}{\rotatebox{90}{Image}}
      & Lines                &    6\,161 &     920 &     929 &     8\,010 \\
      & Running words        &   53\,765 &  8\,599 &  8\,315 &    70\,679 \\
      & Lexicon Size         &    7\,771 &  2\,450 &  2\,492 &     9\,749 \\
      & Character set size   &    72     &      69 &      65 &         81 \\
      \midrule
      \multirow{3}{*}{\rotatebox{90}{Text~}}
      & Running words        &3\,128\,155&  8\,599 &  8\,315 & \!\!3\,145\,069\\
      & Lexicon size         &   19\,892 &  2\,450 &  2\,492 &    20\,773 \\
      & Running OOV(\%)      &        -- &    6.12 &    6.27 &         -- \\
      \midrule
      \multicolumn{2}{c}{WER\,(\%)} & -- &    14.22 &   16.05 &         -- \\
      \bottomrule[0.1em]
    \end{tabular}}
\vspace{-0.5em}
\end{table}

\begin{table}[htbp]
\vspace{-0.5em}
  \centering
  \tabcolsep=4pt
  \caption{
    \label{tab:corpora-others}%
    Basic statistics and partition details of the PAR and CW corpora.
  }
  \vspace{0.2em}
  {
    \renewcommand{\arraystretch}{1.1}
    \begin{tabular}{l|rrrr|rrrr}\cmidrule[0.1em]{1-9}
      & \multicolumn{4}{c|}{PAR} & \multicolumn{4}{c}{GW} \\[-.0em]
       Number of:
      & Training & Valid. & Test   & Total
      & Training & Valid. & Test   & Total  \\
      \midrule
      Lines
      &  2\,237 &     912 &  1\,328 &   4\,477
      &     328 &     164 &     164 &      656 \\
      Running words
      & 14\,042 &  5\,671 &  8\,407 &  28\,120
      &  2\,447 &  1\,224 &  1\,224 &   4\,894 \\
      Lexicon size
      &  3\,221 &  1\,753 &  2\,305 &   4\,936
      &     899 &     539 &     539 &     1471 \\
      Character set size
      &      90 &      80 &      82 &       96
      &      79 &      72 &      72 &       83 \\
      Running OOV(\%)
      &      -- &   14.58 &   12.40 &       --
      &      -- &   29.26 &   24.62 &       -- \\
      \midrule
      WER(\%)
      &      -- &   22.08 &   18.44 &       --
      &      -- &    4.27 &   29.93 &       -- \\
      \bottomrule[0.1em]
    \end{tabular}}
\vspace{-0.3em}
\end{table}

\subsection{Query Sets}\label{sec:query-sets}
\vspace{-0.6em}

Several criteria can be assumed to select the keywords to be used in
KWS assessment experiments.  Clearly, any given KWS system may perform
better or worse depending on the query words it is tested with and how
these words are distributed in the test set.  Of course, the larger
the set of keywords, the more reliable the empirical results.
Moreover, since our approach is aimed at indexing applications,
testing with a large set of keywords is mandatory.

In this work, taking into account these observations, the adopted
criterion for selecting queries for each dataset is to take all the
words that appear in their corresponding training partitions.
%
This is exactly true for BENTHAM, PAR and GW datasets, being for this
latter selected iteratively for each training partition of the 4-fold
cross-validation (Table\,\ref{tab:selection} shows the query set
average size on the 4 cross-validation). Following~\cite{Frinken12},
the line image training lexicon (the one labeled with ``image'' in
Table\,\ref{tab:corpora-iamdb}) excluding punctuation marks and
\emph{stop words} was used as query set for IAM dataset.
Finally, the query set for PLANTAS was extracted from the extended
diplomatic lexicon reported in Table\,\ref{table:PLANTAS_stat},
by previously filtering out
$113$ 
words containing numbers.

It is important to remark that, according to this criterion, there
will be many keywords which do not actually appear in any of the test
images. We say that these keywords are \emph{non-relevant}, while the
remaining ones are \emph{relevant}.
%
Trying to spot non-relevant words is challenging, since other relevant
similar words may be erroneously spotted, which may lead to important
precision degradations.
AP can very aptly used to measure KWS performance for mixed
relevant/non-relevant query sets, but mAP is completely inadequate
because it can only be computed for relevant words
(cf. Sec.\,\ref{sec:evalMeas}).
%
%
%
Table\,\ref{tab:selection} shows the sizes of the query sets used in
the five datasets considered.

\begin{table}[htbp]
\vspace{-1em}
  \centering
  \caption{%
    \label{tab:selection}%
    Sizes of the query sets selected for PLANTAS, BENTHAM, IAM, PAR and GW.
    The amounts of query words for which there is at least one relevant
    test image are also reported.
  }
  \vspace{.5em}                                         
  \tabcolsep=5pt
  \extrarowheight=1pt
  \begin{tabular}{l|rrr}\toprule[0.1em]
    Dataset    & Keywords& Relevant \\
    \midrule
    PLANTAS    & 10\,932 & 4\,888 \\
    BENTHAM    &  8\,658 & 1\,487 \\
    \midrule
    IAM        &  3\,421 & 1\,098 \\
    PAR        &  3\,221 & 1\,218 \\
    GW         &     899 &    234 \\
    \bottomrule[0.1em]
  \end{tabular}
\vspace{-1em}
\end{table}









\subsection{Experimental Setup}                   \label{sec:expSetup}

Following the probabilistic KWS framework proposed in this paper,
experiments have been carried out to assess the capabilities of
specific approaches derived from this framework.  These approaches
require statistical optical character models and language models which
must be trained from the available training images and transcripts.
For language modeling we have just adopted the simple and time-honored
state-of-the-art $N$-gram approach~\cite{Jelinek98}.  But, for optical
modeling, two alternative approaches have been considered: HMMs, which
is a well understood, proper statistical approach and RNNs, which are
recently showing superb performance in HTR.  Both methods have been
used in the main (PLANTAS and BENTHAM) experiments, while only RNNs
have been used in the comparative experiments with IAM, PARZIVAL and
GW.

In general, a similar system architecture is used in all the
experiments.
%
However, depending on the dataset and the optical modelling (HMM or
RNN) adopted, some details of image pre-processing and feature
extraction are different.
The general architecture and these details are discussed the coming
subsections.


\subsubsection*{HMM Optical Modeling}
For HMM optical modelling, line images were preprocessed for slant,
slope and size normalization~\cite{Toselli04,vromero06a} and then
represented as sequences of feature vectors.  Feature extraction (for
PLANTAS and BENTHAM) was based on geometric moments
normalization~\cite{Kozielski:2012}.
%
A left-to-right HMM topology was used for each character, with the
number of states and gaussian densities per state set up taking into
account
character widths an other general dataset features.  Final values of
these two meta-parameters were optimized on the validation partition
of each corpus. More details about these and other meta-parameter
settings for each corpus are given in~\cite{toselli15a,Marti:01}.
%
HMM training 
was carried out using \emph{embedded Baum Welch}
algorithm~\cite{Jelinek98}
using all the training line images and their corresponding
transcripts.

Regarding PLANTAS dataset, note that HMM training (and also RNN
training in the next section) was carried out using the tokenized
diplomatic transcripts, whose basic statistics are presented in
Tab.\,\ref{table:PLANTAS_stat}.

\subsubsection*{RNN Optical Modeling}

Following the recent success of
recurrent neural networks (RNN), in the HTR and KWS fields,
RNN optical modelling was adopted, in addition to HMMs,
to assess the impact of using different probabilistic models on
the proposed KWS framework.

\begin{figure*}[ht]
\centering
\includegraphics[scale=1]{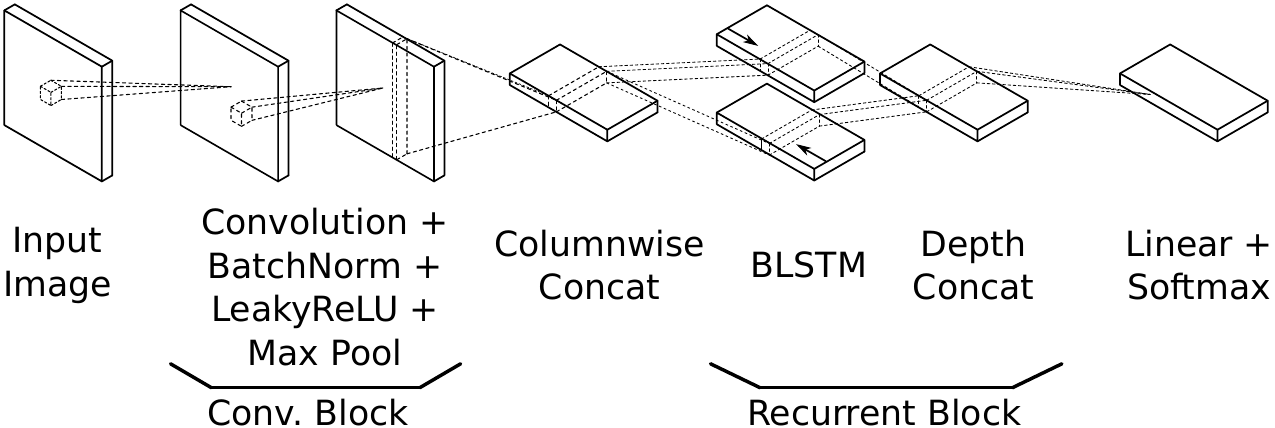}
\caption{Overview of the neural network architecture used in this work.
\label{fig:architecture}}
\end{figure*}

The network is composed of a sequence of four %
\emph{convolution blocks} aimed to extract meaningful features for the
handwritten modeling.  Each of this blocks contains a 2D convolutions
layer~\cite{LeCun:1990} (the number of features extracted in each
block is, respectively, 16, 16, 32 and 32), a batch normalization
layer, a \emph{LeakyReLU}~\cite{Maas:2013} layer as a non-linear
activation function and, finally a $2 \times 2$ \emph{max pooling}
operation is performed (only in the first three layers), in order to
reduce the resolution of the images.
The ``image'' (representation of the extracted features) which results
after these convolution blocks, is processed column-wise by a stack of
three \emph{bidirectional long short-term memory} (BLSTM) recurrent
layers~\cite{Schuster:1997,Graves:2005}. Finally, each column is
linearly transformed to have as many features as characters are in the
particular dataset, plus an additional symbol used by CTC.
A \emph{softmax} transformation is used to interpret the outputs of the
neural network as posterior probabilities.  Dropout is used to reduce
overfitting in between the BLSTM layers and before the final linear
layer output.
A general overview of the architecture is depicted in Figure
\ref{fig:architecture}.

Training is performed using the connectionist temporal classification
(CTC) algorithm~\cite{Graves:2006}, in order to maximize the
(log)probability of the training transcripts, given the corresponding
images.  The parameters of the model are updated using \emph{RmsProp}
\cite{Tieleman:2012}, with a fixed learning rate of 0.0005.

The same architecture and training procedure were used in all the
datasets, except IAM, where
slightly different hyperparameters were used. In this case, we used
the models trained in \cite{Puigcerver:2017}, which have been made
public%
\footnote{\url{https://github.com/jpuigcerver/Laia}}, %
and have shown an excellent performance in HTR tasks~\cite{Puigcerver:2017}.

Finally, in order to combine the output of the neural network with
$N$-gram language models, the posterior probabilities are transformed
into pseudo-likelihoods, using the following equation:
\begin{equation}\label{eq:pseudolikelihoods}
p(\mathbf{x}_t \mid l_t) =
\frac{P(l_t \mid \mathbf{x}_t) \cdot p(\mathbf{x}_t)}
     {P(l_t)}
\approx
\frac{P(l_t \mid \mathbf{x}_t)}{P(l_t)^\gamma}
\end{equation}
where $P(l_t \mid \mathbf{x}_t)$ is the output of the neural network
at column $t$, $P(l_t)$ was estimated by forced alignment on the
training data, and the scaling parameter was fixed to $\gamma=0.2$
for all datasets.

\subsubsection*{Lexicon and Language Modeling}

For each data set, a lexicon was extracted from the corresponding
training partition.
For the benchmarking datasets, the standard tokenization for each
dataset was adopted.  The lexicon of BENTHAM
%
%
was extracted through a specific, rather straightforwardly improved
tokenization scheme, described in detail in~\cite{toselli15a}.  The
lexicon of PLANTAS was similarly extracted from the training
diplomatic transcripts.
For each corpus, except PLANTAS, a $2$-gram language model was
straightforwardly trained from the corresponding training text, using
Kneser-Ney back-off smoothing~\cite{Kneser-Ney:95}.
For PLANTAS, a $2$-gram language model was trained from the raw tagged
training text, as discussed in Sec.\,\ref{sec:datasets}.
Then tags and modernized word versions were removed, from the resulting
model, thereby leaving a diplomatic-only language model for its use in
KWS.
Viterbi decoding meta-parameters associated with each $N$-gram
language model (\emph{grammar scale factor} and %
\emph{word insertion penalty}) were tuned to optimize the WER on the
corresponding validation set.

\subsubsection*{Line Image Word-Graphs}

The KWS approaches used in the experiments rely on line image WGs,
either to compute posteriorgrams, as discussed in
Sec.\,\ref{sec:approaches}, or just to straightforwardly obtain 1-best
image transcription hypotheses.
%
%
WGs are obtained using previously trained optical and language models.
For HMM optical models, the procedure is a well-known variation of
standard Viterbi decoding, as discussed in detail in~\cite{toselli16}.
In the case of RNN, Eq.\,\eqref{eq:pseudolikelihoods} is used to
convert the RNN output character posterior probabilities into
pseudo-likelihoods, which are then considered as emission probabilities
of (typically single-state) character HMMs (see details
in~\cite{Bluche:15}) .  Using these simplified HMMs, the standard HMM
approach is followed both to combine character likelihoods and
language model probabilities and to obtain the WGs as a byproduct of
Viterbi decoding (see~\cite{Kaldi:11,Povey:12}).

As discussed in~\cite{toselli16,toselli17a}, the computational cost of
obtaining a WG grows very fast with the WG size.
Following~\cite{toselli17a}, in our experiments, WG sizes were
controlled by specifying the maximum node input degree~\cite{Young:97}
and/or by applying %
\emph{beam-search pruning} during the HTR decoding process.
In addition, explicit WG pruning~\cite{Zens:2005} was applied to WG
derived from HMM optical models.  These WGs were generally (much)
larger than those derived from RNN optical models.  Therefore, HMM
WGs were explicitly pruned down to sizes which, on the average, were
similar to those of RNN WGs (see Table\,\ref{tab:PLANTAS-BENTHAM-res}).
%


Once WGs had been obtained, they were normalized as discussed
in~\cite{toselli16}). A normalization parameter (called ``logarithm
base factor'' in~\cite{toselli16}) is used in this step to empirically
tune the posterior probability calibration~\cite{Niculescu:05}.  As
with the other meta-parameters, this factor was tuned for each
optical modelling approach on the validation partition of each
corpora.

\subsubsection*{Posteriorgrams and Line-Region Relevance Probabilities}
\label{sec:exp-set-ps-rp}

From the normalized WGs the 1-D posteriorgrams (frame-level word
posterior probabilities, $P(v\mid x,i)$), were obtained as explained
in Sec.\,\ref{sec:wgpost}.
%


Finally, line-level word confidence scores, $P(\RR\mid x,v)$, were
calculated as explained in Sec.\,\ref{sec:regionYNprobs}
and~\ref{sec:YNfromPgram}.  In particular, Eq.\,\eqref{eq:ynFromPgram}
(or, more specifically, its 1-D version,
Eq.\,\eqref{eq:ynFrom1Dpgram}) is used in all the experiments, while
other approaches are tested only in Sec.\,\ref{sec:diffApproxRP}.

In two of these approaches, namely those given by
Eqs.\,\eqref{eq:sumApprox} and\,\eqref{eq:orprobsNB}, a threshold
parameter is employed to find significant local maxima on the 1-D
posteriorgram.
A local maximum is considered as such when the difference %
$P(v\mid x,i)-P(v\mid x,i-1)$
(proportional to the tangent) becomes negative, surpassing in absolute
value a given threshold.  As with the other meta-parameters, this
threshold value is also optimized using validation partitions.


\section{Results}                                  \label{sec:results}

In this section we first use the BENTHAM dataset to empirically
explore the relative performance of the proposed approximations to
compute keyword relevance probabilities.  Then we report final KWS
evaluation results for one specific approximation on the five
datasets before introduced.




\subsection{%
  Testing different Approximations to the Relevance Probability}
\label{sec:diffApproxRP}

A first series of experiments were conducted on the BENTHAM dataset,
using RNN optical modelling, in order to empirically assess and compare
the different approximations for computing the relevance probability
$P(\RR\mid \x,v)$ proposed in Secs.\,\ref{sec:regionYNprobs}
and~\ref{sec:YNfromPgram}.
Table\,\ref{tab:BENTHAM-difAppr} reports the KWS (interpolated
precision) AP performance achieved by these approximations.
%
These approximations range from the roughest one
(Eq.\,\eqref{eq:wordpost}), using the $P(v\mid \x)$, %
to the potentially most accurate, but also much more computationally
expensive approximation, given by by Eq.\,\eqref{eq:relProbMain}.

\begin{table}[htbp]
  \centering
  \caption{BENTHAM AP for different approximations to the relevance
    probability $P(\RR\mid \x,v)$, using RNN optical models.
    $P_{kv\x}$ denotes $\max_{(i,j)\in B_k}\!P(v\mid\x,i,j)$ (see
    Eq.\,\eqref{eq:maxPkvx} in Sec.\,\ref{sec:YNfromPgram}).
    AP$_r$ and mAP$_r$ are reported for the reduced query set of
    $1\,487$ relevant queries (see Table\,\ref{tab:selection}).}
  \label{tab:BENTHAM-difAppr}
  \vspace{.5em}\small
  \begin{tabular}{lr|rr}
    \toprule[0.1em]
    Alternative approximations to $P(\RR\mid \x,v)$
                       &          AP~~~~  &AP$_r$~~~ & mAP$_r$~~ \\
    \midrule[0.1em]
    Eq.\,\eqref{eq:wordpost}: $P(v\mid \x)$
                       &          0.78165 &  0.88407 & 0.94208 \\ 
    \midrule
    Eq.\,\eqref{eq:1bOrApprox}:
     $1\;\text{if}\;v\in\hat{w}$ (1-best transcript); and $0$ otherwise
                       &          0.76297 &  0.82063 & 0.92365 \\ 
    \midrule
    Eq.\,\eqref{eq:sumApprox}: $\sum_{1\leq k\leq n}P_{kv\x}$
                       &          0.87898 &  0.91767 & 0.95438 \\ 
    \midrule
    Eq.\,\eqref{eq:maxApprox} or \eqref{eq:ynFromPgram}:
    $\max_{1\leq k\leq n}P_{kv\x} = \max_{i,j}\!P(v\mid\x,i,j)$
                       &          0.91423 &  0.95155 & 0.95495 \\ 
    \midrule
    \multirow{2}{*}{Eq.\,\eqref{eq:orprobsNB}:
    $\textstyle
      \sum_{k=1}^n P_{kv\x}-\sum_{l<k} P_{kv\x}\,P_{lv\x}
      +
      ~\dots~ (-1)^{n-1} P_{1v\x}\dots P_{nv\x}$}
                       & $\dagger$0.91441 &  0.94888 & 0.95445 \\ 
                       &          0.91289 & $\ddagger$0.95000 & 0.95511 \\
    \midrule
    Eq.\,\eqref{eq:relProbMain}:
     by using forward algorithm on line-region WGs~\cite{toselli16b}
                       &          0.91297 &  0.95023  & 0.95513 \\ 
    \bottomrule[0.1em]
  \end{tabular}
\end{table}

The differences between some of these results are very small and it is
unclear whether small superiorities may be due to a better
approximation accuracies, or just to differences caused by the presence
of non-relevant keywords included in the query set (see
Table~\ref{tab:selection}).
To help better understanding these small differences, the same
experiments were carried out with a smaller query set restricted to
just the $1\,487$ relevant keywords of BENTHAM).  For this smaller
subset the mAP can be also computed; so,
Table\,\ref{tab:BENTHAM-difAppr} also reports the mAP for this subset.
These results are denoted as AP$_r$ and mAP$_r$.
In the case of Eq.\,\eqref{eq:orprobsNB}, the parameter used to
obtain AP and AP$_r$ was optimized on the BENTHAM validation set,
using respectively the whole query set ($\dagger$) and the relevant
query subset ($\ddagger$).  As commented in Sec.\,\ref{sec:evalMeas},
here we also report mAP$_r$ figure only for the relevant query
subset.

AP (and mAP) results for approximations given by
Eqs.\,\eqref{eq:ynFromPgram}, \eqref{eq:orprobsNB}
and\,\eqref{eq:relProbMain} are practically identical, but the first
two are simpler and much less computationally demanding.
The computation needed for Eq.\,\eqref{eq:relProbMain} is very similar
to that of the forward algorithm proposed in Section~IV
of~\cite{toselli16b}, where the very high cost of that algorithm was
empirically studied.
%
%
The results achieved with the other approaches
(Eqs.\,\eqref{eq:wordpost}, \eqref{eq:1bOrApprox}
and~\eqref{eq:sumApprox}) are significantly worse, the naive
1-best KWS technique providing the worst KWS performance.

To summarize, among the approaches considered,
Eq.\,\eqref{eq:ynFromPgram} is as good as the best ones, and also the
fastest and simplest one.  And, also interestingly, it does not have
any meta-parameter which needs to be tuned.  In what follows, all the
results will be reported only for this approach.

\vspace{0.5em}
\subsection{Comparing the Impact of Using
                               HMM or RNN Optical Models} \label{sec:expLarge}
\vspace{-0.5em}

Using only Eq.\,\eqref{eq:ynFromPgram} (or its 1-D version,
Eq.\,\eqref{eq:ynFrom1Dpgram}), the second series of experiments were
devoted to study how different optical modelling choices (HMM and
RNN), adopted to compute the posteriorgrams, affect the KWS
performance.  In addition to the BENTHAM dataset, the other large
dataset presented in Sec.\,\ref{sec:datasets} (PLANTAS) is considered.
Table\,\ref{tab:PLANTAS-BENTHAM-res} shows the results of this study.

%
%

\begin{table}[htbp]
\vspace{-0.5em}
  \centering
  \caption{KWS performance for PLANTAS and BENTHAM
    using the two optical modelling approaches: HMM and RNN.
    \label{tab:PLANTAS-BENTHAM-res}}
  \vspace{.5em}
  \begin{tabular}{llcr}
    \toprule[0.1em]
    Dataset & Optical models &   AP  & \#Spots/Line \\
    \midrule[0.1em]
    \multirow{5}{*}{PLANTAS}
    & HMMs $1$-best          & 0.722 & 9.7~~~~ \\
    & HMMs                   & 0.912 & 311.9~~~~ \\
    & HMMs (pruned-WGs)      & 0.909 & 128.2~~~~ \\
    \cmidrule{2-4}
    & RNN                    & 0.924 & 133.9~~~~ \\
    & RNN $1$-best           & 0.794 & 9.7~~~~ \\
    \midrule[0.1em]
    \multirow{5}{*}{BENTHAM}
    & HMMs $1$-best          & 0.740 & 8.9~~~~ \\
    & HMMs                   & 0.910 & 335.8~~~~ \\
    & HMMs (pruned-WGs)      & 0.907 &  71.2~~~~ \\
    \cmidrule{2-4}
    & RNN                    & 0.914 &  72.2~~~~ \\
    & RNN $1$-best           & 0.763 &  8.4~~~~ \\
    \bottomrule[0.1em]
  \end{tabular}
\end{table}

For completeness, results obtained with the naive $1$-best KWS approach
(Eq.\,\eqref{eq:1bOrApprox}) are also given.
%
For HMMs, results using both original and explicitly pruned WGs are
reported.  As discussed in Sec.\,\ref{sec:expSetup}, this pruning
produces WGs of similar average size as that of WGs obtained with RNN
optical models.
Rather than showing the specific sizes of the WGs (which are
meaningless and can be safely discarded once the relevance
probabilities are computed), the average number of the spotting
alternatives produced for each line region (``Spots/Line'') are
reported.  For indexing applications, the overall memory requirements
are proportional to this ``spotting density''.
As observed in Tab\,\ref{tab:PLANTAS-BENTHAM-res}, the AP performance
for the explicitly pruned WGs is not significantly degraded.


Recall\,--\,\,interpolated-Precision (R-P) curves for PLANTAS and
BENTHAM obtained using both HMM and RNN optical modelling are plotted
in Fig.\,\ref{fig:R-P_PLANTA-BENTHAM}.

\begin{figure}[htbp]
  \centering
  \includegraphics[width=.45\textwidth]{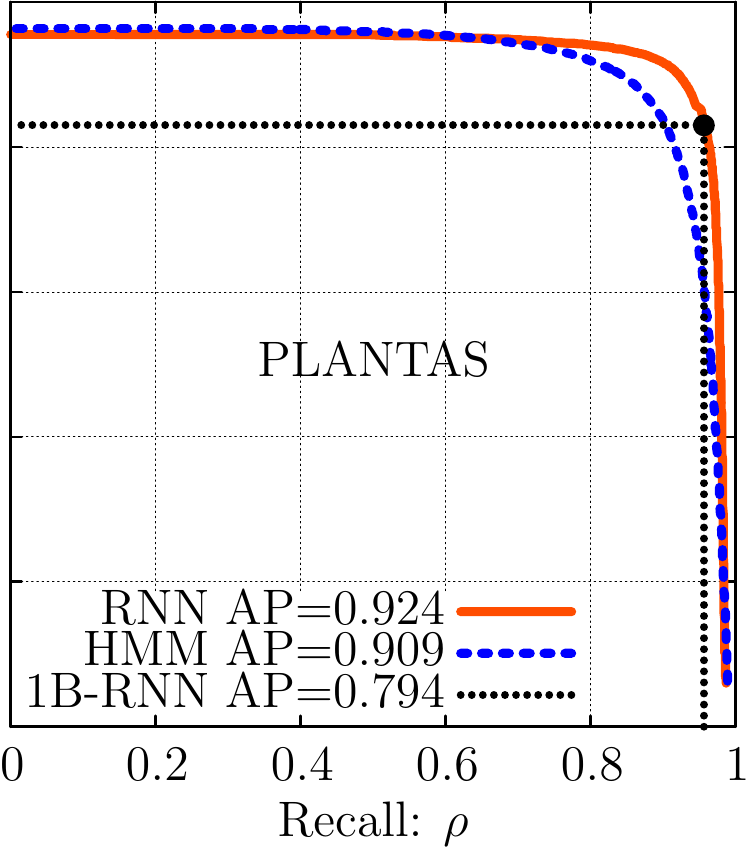}%
  \hspace{2em}%
  \includegraphics[width=.45\textwidth]{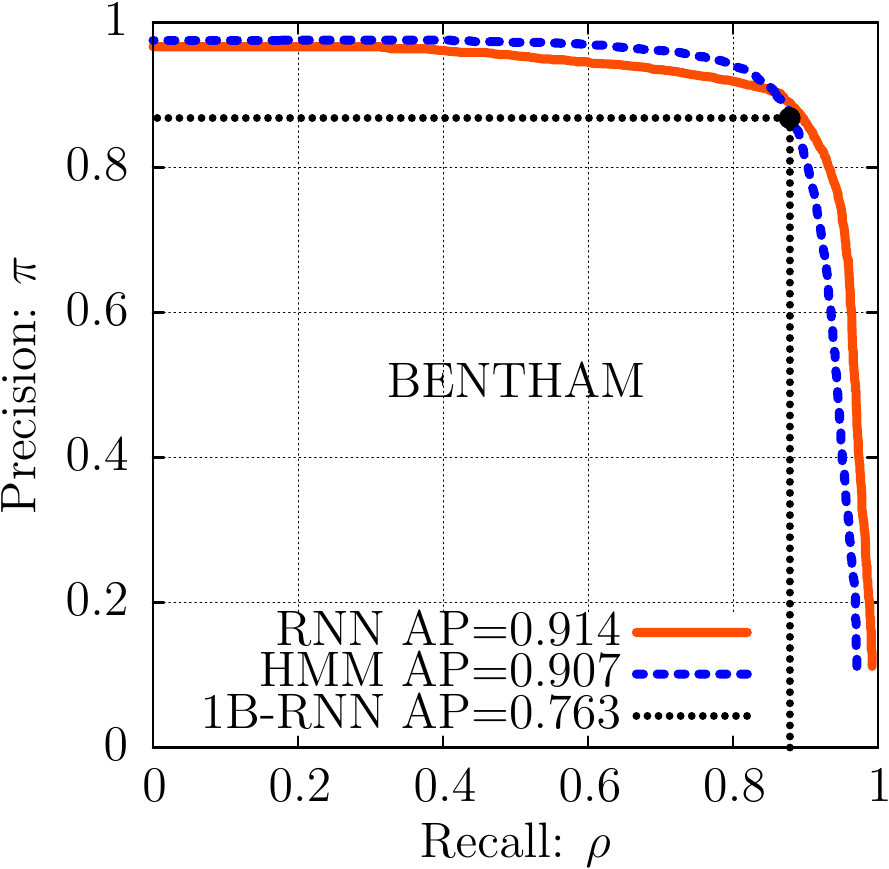}%
  \caption{\label{fig:R-P_PLANTA-BENTHAM}%
    Interpolated R-P curves for PLANTAS and BENTHAM, using optical
    modelling based on HMMs and RNN.  The (degenerate) curves for naive
    1-best KWS are also shown}
\end{figure}

According to the results in this section, only small KWS performance
differences are observed depending on the choice of HMM or RNN for
optical modelling.
This applies to both PLANTAS and BENTHAM, both of which are fairly
large handwritten corpora and entail important differences between
them in many handwriting processing aspects.
While RNNs are generally known to be better then HMMs for character
optical modeling, the present results suggest that this superiority
mainly affects to the mode of the modeled distributions -- thereby
generally leading to better character error rate results.  However,
when, as in KWS, the whole distribution is brought into play, the
superiority becomes less obvious.

These observations are in line with those of~\cite{toselli15a}, where
the transcription performance of both the HMM- and RNN-based HTR systems
were also similar, mainly due to the good feature extraction employed
for the HMM modelling approach.

\subsection{Experiments with Common Benchmarking Data Sets}
                                                       \label{sec:expMisc}

Additional KWS experiments were carried out with three well established
benchmark datasets (see Sec.\,\ref{sec:datasets}): \textsc{IAM},
\textsc{Parzival} (PAR) and \textsc{George Washington} (GW).
The R-P curves and AP values obtained for these datasets, using RNN
optical modelling and relevance probability approximation given by
Eq.\,\eqref{eq:maxApprox}, are presented in
Fig.\,\ref{fig:R-P_otherSets}.

\begin{figure*}[htbp]
\vspace{-0.5em}
  \begin{minipage}{.5\textwidth}
    \centering\small
    \includegraphics[width=\textwidth]{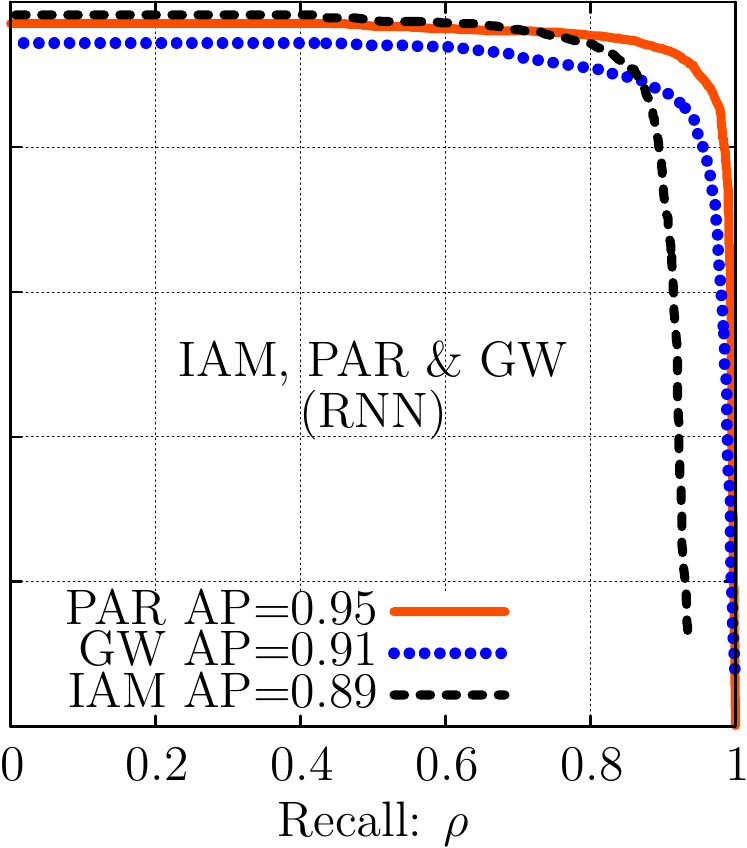}
    \caption{\label{fig:R-P_otherSets}%
      Interpolated R-P curves for the IAM, PAR and GW
      datasets, using RNN optical modelling and the
      relevance probability approximation given by
      Eq.\,\eqref{eq:maxApprox}.}
  \end{minipage}
  \hfill%
  \raisebox{0.0em}{
    \begin{minipage}{.45\textwidth}
      \centering
      \renewcommand{\figurename}{Table}
      \setcounter{figure}{7}
      \caption{%
        AP results achieved by several line-region KWS approaches on
        IAM, PAR and GW, with varied empirical setups which may be
        loosely compared with ours. Values marked with $\dag$ were
        obtained using a different query set. The three best results
        for each dataset are marked in boldface.}
      \label{tab:otherResults}
      \vspace{12pt}
      \scriptsize
      \extrarowheight=0pt
      \tabcolsep=5pt
      \begin{tabular}{c>{\bf}l>{~~~}lll}\toprule[0.1em]
        Ref. &
        ~~~~Approach &
        \hspace{-0em}IAM &
        \hspace{-.0em}PAR &
        GW \\
        \midrule\midrule
        \multirow{1}{*}{\cite{Wicht17,rodriguez2008}} &
        LGHF-HMM &
        0.05 $\dag$ &
        0.25 $\dag$ &
        0.33 $\dag$ \\ \midrule
        \multirow{2}{*}{\cite{Fischer17}} &
        BP-FR &
        ---   &
        0.58 $\dag$ &
        0.56 $\dag$ \\
        &
        BP    &
        ---   &
        0.61 $\dag$ &
        0.55 $\dag$ \\ \midrule
        \multirow{1}{*}{\cite{wicht16}} &
        CDBN-DTW &
        ---   &
        0.59 $\dag$ &
        0.56 $\dag$ \\ \midrule
        \multirow{1}{*}{\cite{kumar14}} &
        BLRC & 0.49 $\dag$ & --- & --- \\ \midrule
        \multirow{2}{*}{\cite{Fischer13}} &
        Classic Filler-HMM &
        0.48 $\dag$ &
        ---  &
        0.72 $\dag$ \\
        &
        $2$-gram Filler-HMM &
        0.55 $\dag$ &
        ---  &
        0.74 $\dag$ \\ \midrule
        \multirow{2}{*}{\cite{Fischer12}} &
        DTW &
        --- &
        0.39 $\dag$ &
        0.44 $\dag$ \\
        &
        Classic Filler-HMM       &
        --- &
        0.86 $\dag$ &
        0.62 $\dag$ \\ \midrule[0.1em]
        \multirow{2}{*}{\cite{Wicht17}} &
        CDBN-DTW &
        0.01 $\dag$ &
        0.73 $\dag$ &
        0.57 $\dag$ \\
        &
        CDBN-HMM &
        0.65 $\dag$ &
        \textbf{0.92} $\dag$ &
        0.71 $\dag$ \\ \midrule
        \multirow{1}{*}{\cite{Wicht17,Terasawa09}} &
        HOG-HMM &
        0.65 $\dag$ &
        \textbf{0.92} $\dag$ &
        0.68 $\dag$ \\ \midrule
        \multirow{1}{*}{\cite{Terasawa09}} &
        HOG-DTW &
        --- &
        --- &
        0.79 $\dag$ \\ \midrule
        \multirow{1}{*}{\cite{Wshah:14}} &
        Filler-B & 0.58 $\dag$ & --- & --- \\ \midrule
        \multirow{1}{*}{\cite{toselli16b}} &
        $6$-gram CL-Filler &
        0.61 &
        --- &
        --- \\ \midrule
        \multirow{1}{*}{\cite{toselli16}} &
        HMM + $2$-gram LM &
        \textbf{0.72} &
        0.89 &
        \textbf{0.77} \\ \midrule
        \multirow{3}{*}{\cite{Frinken12}} &
        DTW  &
        0.02 &
        0.37 &
        0.48 \\
        &
        Classic Filler-HMM  &
        0.36 &
        0.83 &
        0.60 \\
        &
        BLSTM  &
        \textbf{0.78}   &
        \textbf{0.94}   &
        \textbf{0.84}   \\
      \midrule \midrule
        \multicolumn{2}{c}{\textbf{Proposed approach}} &
        \textbf{0.89} &
        \textbf{0.95} &
        \textbf{0.91} \\
        \bottomrule[0.1em]
        \multicolumn{3}{c}{} \\[-.5em]
      \end{tabular}
    \end{minipage}}
\end{figure*}

To place our results in comparison with previously published work,
Table\,\ref{tab:otherResults} presents (word) segmentation-free,
query-by-string KWS results obtained by other authors on the same
three datasets.
The following approaches have been considered: our previous work on
lexicon-based HMM KWS~\cite{toselli16}, bayesian logistic regression
classifier (BLRC)~\cite{kumar14}, character-lattice-based KWS
(CL-based)~\cite{Toselli13a,toselli16b}, dynamic time warping
(DTW)~\cite{wicht16,Wicht17,Frinken12,Fischer12},
bipartite graph matching without rejection (BP) and with fast
rejection (BP-FR)~\cite{Fischer17}, bidirectional long-short term
memory (BLSTM)~\cite{Frinken12}, convolutional deep belief network
(CDBN)~\cite{wicht16}, local gradient histogram features
(LGHF)~\cite{Wicht17,rodriguez2008}, HMM-filler with background
modelling (Filler-B)~\cite{Wshah:14} and histogram of gradients
(HOG)~\cite{Wicht17,Terasawa09}.
The references with two cites like \cite{Wicht17,rodriguez2008}
and~\cite{Wicht17,Terasawa09} indicate that the methods originally
presented in~\cite{rodriguez2008} and~\cite{Terasawa09} have been
applied to obtain comparative results under the experimental setup
described in~\cite{Wicht17}.
The pyramidal histogram of characters (PHOC) KWS approach presented
in~\cite{almazan14} is \emph{not} included because it is
\emph{segmentation-based}~\cite{Giotis:2017}%
\footnote{In~\cite{almazan14} PHOC results are misleadingly compared
  with, and considered superior to other state-of-the-art
  \emph{segmentation-free} methods such as BLSTM~\cite{Frinken12}}.

It should be noted that the experimental setups adopted in some of
these works may vary significantly with respect to the setup adopted
in this work.  In particular, in the entries
marked with~$\dag$, KWS performance was obtained using a query set
selected from the test partition.  Also, in some cases it is not
completely clear whether the results are provided in terms of AP or
mAP.
Therefore, the results of Table\,\ref{tab:otherResults} can only be
considered loosely comparable.

Notwithstanding the slight differences, we conclude that the
superiority of the methods proposed in this work can be acknowledged.

%
%

\section{Conclusions and Outlook}            \label{sec:conclusion}
\vspace{-0.5em}

A probabilistic framework for query-by-string,
(word-)segmentation-free, lexicon-based KWS, aimed at indexing the
textual contents of large collections of handwritten text images, has
been presented.
The formulation of this framework makes it self-evident that KWS is
always implicitly or explicitly based on word recognition posterior
probabilities and provides probabilistic interpretations to
many classical KWS views and methods.
Various developments of this framework into specific KWS approaches
have been proposed and empirically evaluated.  The most efficient and
effective of these approaches are based on a word-graph representation
of the joint probability distribution of a (line) image region and the
text contained in this region.
As discussed in previous works, this joint distribution can be
advantageously obtained using the same statistical models and training
methods, and similar decoding procedures as those used for modern,
segmentation-free handwritten text recognition approaches.  Therefore,
our approaches follow this idea.
According to empirical results achieved on several traditional
datasets and new, larger corpora, the proposed approaches outperform
all the methods proposed and tested so far.


The KWS approaches presented in this paper are all
\emph{lexicon-based} (LB).  As applied to large-scale indexing, LB
methods in general are known to be faster and more accurate than
\emph{lexicon-free} (LF) ones, based on raw character processing.
However, since LB KWS relies on a predefined lexicon, fixed in the
training phase, it does not support queries involving
out-of-vocabulary (OOV) keywords.
A basic lexicon can be straightforwardly derived from the training
transcripts and it can be expanded by including other words expected
to appear in the handwritten image collection being considered.  These
words can be derived from similar texts or glossaries and/or from
available vocabularies of the same language and historical period.
These issues have not been sufficiently studied in this work.  In
fact, aiming at obtaining results comparable with other
state-of-the-art KWS approaches, the experiments were carried out with
query sets extracted from the training texts, thereby guaranteeing
that all query words are in-vocabulary.
It is worth noting, however, that while the OOV problem may be serious
if the indexed vocabulary is small, it becomes much less important
with very large vocabularies -- which is generally the case in real
indexing applications.  Several live demonstrators which support this
fact can be tested at the Demonstrations page of the
\textsc{tranScriptorium} project web site.\!\footnote{%
  \url{http://transcriptorium.eu/demonstrations} $\rightarrow$
  Keyword Indexing \& Search} %
These systems also include flexible queries such as searching for word
sequences and the BOOLEAN AND/OR/NOT query methods described
in~\cite{Noya:17}.

In any case, we are currently developing new approaches for full LF
indexing, without having to sacrifice the high effectiveness and
efficiency of LB techniques.  Two families of methods are being
studied.

The first one capitalizes on the very good performance for
in-vocabulary keywords of LB methods based on the here proposed
probabilistic KWS framework and solves the problem at the query
phase, only for OOV words.  The idea is to smooth the (implicitly
null) relevance probabilities of OOV keywords by relying on the
indexed probabilities of ``similar'' in-vocabulary words.  Most of our
work in this direction is reviewed or presented
in~\cite{Puigcerver:17}.
While reasonably good results are achieved with these methods, they
always entail query response time penalties for OOV queries -- and
these penalties can become prohibitive for large collections of say
hundreds of thousands or millions of images.


The other family, also within the here proposed probabilistic KWS
framework, abandons the use a lexicon altogether and favors working
at the character level.  However, it also attempts to keep the good
performance of LB indexing by producing relevance probabilities for
``words'' (actually arbitrary character sequences) which are
``discovering'' in the very test images (those to be indexed).  This
kind of work is now in progress, but some of the key ideas have
already been published in~\cite{Toselli:15d,toselli16b,Toselli:2017c}.
Moreover, the large-scale (75\,000 page images) indexing task
described in~\cite{Toselli:2017c} has recently been fully accomplished
and the resulting real search system can be used from the publicly
available HIMANIS project search interface.\!\footnote{%
  \url{http://prhlt-kws.prhlt.upv.es/himanis}%
} %
The techniques used in this system will be presented in an upcoming
publication devoted to probabilistic LF KWS.

%

%

\section{Acknowledgments}

This work has been partially supported through

%



\bibliographystyle{abbrv}

\end{document}